\documentclass[8pt]{revtex4}
\usepackage{amsmath}
\usepackage{amsfonts}
\usepackage{amssymb}
\usepackage{latexsym}
\usepackage{epsfig}

\begin{document}

\title{  The evolution of Brown-York quasilocal energy due to evolution of Lovelock gravity in a system of M0-branes}

\author{Alireza Sepehri $^{1,2}$}
\email{alireza.sepehri@uk.ac.ir} \affiliation{ $^{1}$Faculty of
Physics, Shahid Bahonar University, P.O. Box 76175, Kerman,
Iran.\\$^{2}$ Research Institute for Astronomy and Astrophysics of
Maragha (RIAAM), P.O. Box 55134-441, Maragha, Iran. }

\author{Farook Rahaman}
\email{rahaman@associates.iucaa.in} \affiliation{Department of
Mathematics, Jadavpur University, Kolkata 700 032, West Bengal,
India.}

\author{Salvatore Capozziello $^{4,5,6,7}$}
\email{capozziello@na.infn.it} \affiliation{$^{4}$ Dipartimento di
Fisica, Universit´a di Napoli ”Federico II”, I-80126 - Napoli,
Italy. \\
$^{ 5 }$ INFN Sez. di Napoli, Compl. Univ. di Monte S. Angelo,
Edificio G, I-80126 - Napoli, Italy.\\ $^{  6}$ Gran Sasso Science
Institute (INFN), Viale F. Crispi, 7, I-67100, L'’Aquila, Italy.\\
$^{7}$Tomsk State Pedagogical University, ul. Kievskaya, 60, 634061 Tomsk, Russia.}

\author{Ahmed Farag Ali}
\email{ahmed.ali@fsu.edu.eg} \affiliation{Deptartment of Physics,
Faculty of Science, Benha University, Benha 13518, Egypt.}

\author{Anirudh Pradhan}
\email{pradhan@associates.iucaa.in} \affiliation{Department of
Mathematics, Institute of Applied Sciences and Humanities, G L A
University, Mathura-281 406, Uttar Pradesh, India}

\begin{abstract}
Recently, it has been suggested in [JHEP 12(2015)003] that the Brown-York 
mechanism can be used to measure the quasilocal energy in Lovelock gravity.
We have used this method in a system of M0-branes and show that
the Brown-York energy evolves in the process of birth and growth
of Lovelock gravity. This can help us to predict phenomenological
events which are emerged as due to dynamical structure of Lovelock
gravity in our universe. In this model, first, M0-branes join to
each other and form an M3-brane and an anti-M3-branes connected by
an M2-brane. This system is named BIon. Universes and
anti-universes live on M3-branes and M2 plays the role of wormhole
between them. By passing time, M2 dissolves in M3's and nonlinear
massive gravities, like Lovelock massive gravity, emerges and grows.
By closing M3-branes, BIon evolves and wormhole between branes
makes a transition to black hole. During this stage, Brown-York
energy increases and shrinks to large values at the colliding
points of branes. By approaching M3-branes towards each other, the
square energy of their system becomes negative and some tachyonic
states are produced. To remove these states, M3-branes compact,
the sign of compacted gravity changes, anti-gravity is created
which leads to getting away of branes from each other. Also, the
Lovelock gravity
disappears and it's energy forms a new M2 between M3-branes. 
By getting away of branes from each other, Brown-York energy 
decreases and shrinks to zero. \\\\


 \end{abstract}
 \date{\today}

%%%%%%%%%%%%%%%%%%%%%%%%%%%%%%%%%%%%%%%%%%%%%%%%%%%%%%%% Section 1 %%%%%%%%%%%%%%%%%%%%%%%%%%%%%%

 \maketitle

\section{Introduction}
Measuring the quasilocal energy in general relativity is an
extremely important but a long eluding problem. There is an
ambiguity in calculating this energy as due to its nonlinear
nature and the fact that gravitational energy is non-localizable.
The best method for measuring gravitational energy has been
proposed by the Brown and York in \cite{m1} by employing the
Hamilton-Jacob theory for gravitation. This method has been
studied extensively in general relativity for static spacetime
with spherical symmetry and also for Kerr black hole
\cite{s1,s2,s3,s4}. Recently, It has been generalized the
Brown-York formalism to all orders in Lovelock gravity and have been
verified the conjunction for pure Lovelock black hole \cite{s5}.
Motivated by this research, we investigate the evolution of this
energy during the birth and growth of Lovelock gravity in a system
of M0-branes. In our model, M0-branes join to each other and form
universe and anti-universe which are connected by a wormhole.
This system is named BIon. By considering the changes in
Brown-York energy, we can predict the evolution of Lovelock
gravity in a BIonic system. This can help us predict
phenomenological events which are emerged due to dynamical
structure of Lovelock gravity in our universe.

In previous work, BIon model \cite{m3,m4,m5,m6,m7,m8} has been used in cosmology to consider the evolution of universe from birth and then inflation to late time acceleration. For example, in one paper, k black fundamental strings link to each other and construct a BIon \cite{m3,m4}. Coincidence with the birth of BIon, a universe and an anti-universe are born  which are connected with each other through a wormhole \cite{m3,m4}. This wormhole dissolves in branes and causes their expansion. In another work, first, N fundamental
strings transit to N pair of M0 and anti-M0-branes. Then,
M0-branes glue to each other and build a pair of an M5 and an
anti-M5-brane. This pair decays to an M3-brane, an anti-M3-brane
in additional to one M2-brane. Our universe is formed on one of
these M3-branes and M2 has the role of wormhole . In this theory,
there is not any big-bang  and  the origin of universe is a
fundamental string \cite{m7}. In another scenario, first M0-branes
are compacted on  a circle and N D0-branes are created.  Then,  N
D0-branes join to each other and build one D5-branes.  Next, D5-
brane is compacted on two circles and a D3-brane, two D1-brane and
some extra energies are produced. Our universe is built on D3-brane 
and D1-branes  dissolve in universe and lead to inflation and late time
acceleration \cite{m8}.

Now, a natural question arises to how can we consider the evolution of
BIon in four dimensional universe? We answer it by measuring the
Brown-York energy which changes by changing the Lovelock gravity
in this system. In our consideration, first M0-branes join to each
other and build an M3, an anti-M3 and an M2-brane which connects
them. Each of universes live on an M3-brane and M2 has the role of
wormhole. By passing time, this M2 dissolves into M3-branes, some
nonlinear gravities like the $F(R)$-gravity and Love-lock gravity
\cite{m2,mm2} emerge and universe expands. By closing branes
towards each other, wormhole becomes more thermal and transits to
black hole. During this era, the Brown-York energy increases and
shrinks to large values. By approaching branes to each other, the
square energy of their system becomes negative and some tachyonic
states are produced. To remove these states, M3 compacts, the sign
of compacted gravity changes and the anti-gravity emerges which
leads to getting away of branes from each other and contraction 
of universe. In this epoch, the Brown-York energy decreases and shrinks to zero.\\

The outline of the paper is the  following.  In section \ref{o1},
we consider the evolution of the Brown-York energy during the
expansion branch in BIonic system. In section \ref{o2}, we discuss as to how by compacting branes gravity changes to anti- gravity and the particles get away from each other. This leads to a decrease of  the Brown-York energy in BIonic system. The last section is devoted to discussion and conclusions.

%%%%%%%%%%%%%%%%%%%%%%%%%%%%%%%%%%%%%%%%%%%%%%%%%%%%
\section{ the evolution of the Brown-York energy during the
expansion branch in BIonic system }\label{o1}

In this section, we show that the relevant action of Mp-branes can
be obtained by summing over the actions of p M0-branes. Then, we
obtain the Brown-York energy in a system of M2-M3 branes. In this
system, universes are placed on M3-branes and connect with each
other via an M2-brane. M2 plays the role of wormhole, dissolves in
M3-branes and leads to the emergence of nonlinear gravity and a
black hole. In the background of this gravity, the Brown-York
energy increases and tends to large values. It has been shown that the Brown-York energy can be
obtained from following equation \cite{m1}:
\begin{eqnarray}
E_{BY}=\frac{1}{8\pi} \int_{\beta}d^2x
\sqrt{q}(\tilde{k}-\tilde{k}_{0})\label{p1}
\end{eqnarray}
where q is the metric defined on the  surface $B$ which is the
boundary of three space $\Sigma$ and $\tilde{k}_{0}$ denotes the
trace of extrinsic curvature for some reference spacetime. We will
use of this equation and consider the evolution of energy in BIon.
To begin considering the evolution of energy in BIon, we should
first consider the process of formation of  BIon in M-theory. To
this end, we should construct the action of BIon from D1 and
M1-branes and also show that these branes are built of D0 and
M0-brane. We introduce the Lagrangian for D1 as
\cite{m7,m8,m9,m10,m11,m12,m13,m14}:

\begin{eqnarray}
S = - T_{D1}\int  d^{2}\sigma ~ STr \Bigg(-det(P_{ab}[E_{mn}
E_{mi}(Q^{-1}+\delta)^{ij}E_{jn}])det(Q^{i}_{j})\Bigg)^{1/2}~~
\label{r1}
\end{eqnarray}
where
\begin{eqnarray}
   E_{mn} = G_{mn} + B_{mn}, \qquad  Q^{i}_{j} = \delta^{i}_{j} + i\lambda[X^{j},X^{k}]E_{kj} \label{r2}
\end{eqnarray}
where $\lambda=2\pi l_{s}^{2}$, $G_{ab}=\eta_{ab}+\partial_{a}X^{i}\partial_{b}X^{i}$ and $X^{i}$ 
are scalar fields of mass dimension. Here $a,b=0,1,...,p$
are the world-volume indices of the Dp-branes, $i,j,k = p+1,...,9$
are indices of the transverse space, and m,n are the
ten-dimensional spacetime indices. Also,
$T_{Dp}=\frac{1}{g_{s}(2\pi)^{p}l_{s}^{p+1}}$ is the tension of
Dp-brane, $l_{s}$ is the string length and $g_{s}$ is the string
coupling. To obtain the Lagrangian for Dp-brane, we should use of
following rules \cite{m7,m8}:

\begin{eqnarray}
&& \Sigma_{a=0}^{p}\Sigma_{m=0}^{9}\rightarrow \frac{1}{(2\pi l_{s})^{p}}\int d^{p+1}\sigma \Sigma_{m=p+1}^{9}\Sigma_{a=0}^{p} \qquad \lambda = 2\pi l_{s}^{2}, \nonumber \\
&& i,j=p+1,..,9\qquad a,b=0,1,...p\qquad m,n=0,1,..,9 , \nonumber \\
&& i,j\rightarrow a,b \Rightarrow [X^{a},X^{i}]=i \lambda
\partial_{a}X^{i}\qquad \frac{1}{Q}\rightarrow \sum_{n=1}^{p}
\frac{1}{Q}(\partial_{a}X^{i}\partial_{b}X^{i})^{n},~~~ det(Q^{i}_{j})\rightarrow det(Q^{i}_{j})\prod_{n=1}^{p}
det(\partial_{a_{n}}X^{i}\partial_{b_{n}}X^{i}). \label{r3}
\end{eqnarray}
By substituting above mappings in action (\ref{r1}), we obtain the antisymmetric scheme of the Lagrangian for Dp-brane \cite{m7,m8,m9,m10,m11,m12,m13,m14}:

\begin{eqnarray}
 &&S=-\frac{T_{Dp}}{2}\int d^{p+1}x \sum_{n=1}^{p}\beta_{n} \chi^{\mu_{0}}_{[\mu_{0}}\chi^{\mu_{1}}_{\mu_{1}}...\chi^{\mu_{n}}_{\mu_{n}]}, \label{s4}
\end{eqnarray}
where
\begin{eqnarray}
 &&\chi^{\mu}_{\nu}\equiv
\sqrt{g^{\mu\rho}\partial_{\rho}X^{i}\partial_{\nu}X^{j}\eta_{ij}}\label{s5}
\end{eqnarray}
and $X^{a}$'s are scalars, $\mu,\nu=0,1,...,p$ denote the
world-volume indices of the $Mp$-branes, $i,j,k = p+1,...,9$ refer
to indices of the transverse space and $\beta$ is a constant.
Also, $T_{Dp}=\frac{1}{g_{s}(2\pi)^{p}l_{s}^{p+1}}$ is the tension
of Dp-brane, $l_{s}$ is the string length and $g_{s}$ is the
string coupling. Using the method in ref \cite{m7}, we can write
the following mappings  \cite {m7,m8,m9,m10,m11,m12,m13,m14}:

\begin{eqnarray}
 && [X_{\rho},X^{i}]=i\lambda \partial_{\rho}X^{i}\rightarrow \chi^{\mu}_{\nu}\equiv
\sqrt{g^{\mu\rho}[X_{\rho},X^{i}][X_{\nu},X^{j}]\eta_{ij}}, 
\nonumber\\&& \Sigma_{m=0}^{9} \rightarrow
\Sigma_{a,b=0}^{p}\Sigma_{j=p+1}^{9}\label{s6}
\end{eqnarray}
Substituting mappings of equation (\ref{s6}) in equation
(\ref{s5}), we obtain:
\begin{eqnarray}
&& S_{Dp} = -(T_{D0})^{p} \int dt \sum_{n=1}^{p}\beta_{n}\Big(
\delta^{a_{1},a_{2}...a_{n}}_{b_{1}b_{2}....b_{n}}L^{b_{1}}_{a_{1}}...L^{b_{p}}_{a_{p}}\Big)^{1/2}, \nonumber\\&&
(L)^{b}_{a}=Tr( \Sigma_{a,b=0}^{p}\Sigma_{j=p+1}^{9}
[X^{a},X^{j}][X_{b},X_{j}]). \label{s7}
\end{eqnarray}
where we have used of antisymmetric properties of $\delta$ and definition of the action of D0-brane \cite
{m7,m8,m9,m10,m11,m12,m13,m14}:
\begin{eqnarray}
&& S_{D0} = -T_{D0} \int dt Tr( \Sigma_{m=0}^{9}
[X^{m},X^{n}]^{2}) \label{s8}
\end{eqnarray}
Equation (\ref{s7}) shows that each Dp-brane can be built from joining p D0-branes.  We can extend these results to M-theory. By replacing two dimensional Nambu-Poisson bracket for Dp-branes by three one in action and using the Li-3-algebra \cite{m15,m16,m17,m18}, we can obtain the relevant action for M0-brane \cite {m7,m8}:

\begin{eqnarray}
S_{M0} = T_{M0}\int dt Tr( \Sigma_{M,N,L=0}^{10}
\langle[X^{M},X^{N},X^{L}],[X^{M},X^{N},X^{L}]\rangle) \label{s9}
\end{eqnarray}
where $X^{M}=X^{M}_{\alpha}T^{\alpha}$ and
\begin{eqnarray}
 &&[T^{\alpha}, T^{\beta}, T^{\gamma}]= f^{\alpha \beta \gamma}_{\eta}T^{\eta} \nonumber \\&&\langle T^{\alpha}, T^{\beta} \rangle = h^{\alpha\beta} \nonumber \\&& [X^{M},X^{N},X^{L}]=[X^{M}_{\alpha}T^{\alpha},X^{N}_{\beta}T^{\beta},X^{L}_{\gamma}T^{\gamma}]\nonumber \\&&\langle X^{M},X^{M}\rangle = X^{M}_{\alpha}X^{M}_{\beta}\langle T^{\alpha}, T^{\beta} \rangle
\label{s10}
\end{eqnarray}
where  $X^{M}$(i=1,3,...10) are transverse scalars to M0-brane. Replacing the action of D0 by M0 in the action (\ref{s7}), we get:

\begin{eqnarray}
&&S_{Mp} = -(T_{M0})^{p} \int dt \sum_{n=1}^{p}\beta_{n}\Big(
\delta^{a_{1},a_{2}...a_{n}}_{b_{1}b_{2}....b_{n}}L^{b_{1}}_{a_{1}}...L^{b_{p}}_{a_{p}}\Big)^{1/2},\nonumber\\&&
(L)^{a}_{b}= Tr(  \Sigma_{a,b=0}^{p}\Sigma_{j=p+1}^{10}
\langle[X^{a},X^{i},X^{j}],[X_{b},X_{i},X_{j}]\rangle). \label{s11}
\end{eqnarray}
We can write the more complete form of the action by regarding the non-commutative brackets on the brane:

\begin{eqnarray}
&&S_{Mp} = -(T_{M0})^{p} \int dt \sum_{n=1}^{p}\beta_{n}\Big(
\delta^{a_{1},a_{2}...a_{n}}_{b_{1}b_{2}....b_{n}}L^{b_{1}}_{a_{1}}...L^{b_{p}}_{a_{p}}\Big)^{1/2}\nonumber\\&&
(L)^{a}_{b}= Tr(  \Sigma_{a,b=0}^{p}\Sigma_{j=p+1}^{10}\Big(
\langle[X^{a},X^{i},X^{j}],[X_{b},X_{i},X_{j}]\rangle+
\langle[X^{a},X^{c},X^{d}],[X_{b},X_{c},X_{d}]\rangle+\langle[X^{k},X^{i},X^{j}],[X_{k},X_{i},X_{j}]\rangle
\Big))\nonumber \\  \label{s12}
\end{eqnarray}
This action shows that similar to string theory, each Mp-brane can be constructed from p M0-branes. When M0-branes join to each other, two form gauge fields are born which play the role of tensor mode of graviton on the brane. To show this, we use of  the mechanism in ref \cite{m7} and obtain the following mappings \cite{m7,m8,m15,m16,m17,m18}:

\begin{eqnarray}
&&\langle[X^{a},X^{i},X^{j}],[X^{a},X^{i},X^{j}]\rangle=
\frac{1}{2}\varepsilon^{abc}\varepsilon^{abd}(\partial_{c}X^{i}_{\alpha})(\partial_{d}X^{i}_{\beta})
\langle(T^{\alpha},T^{\beta})\rangle \sum_{j} (X^{j})^{2} =
 \frac{1}{2}\langle \partial_{a}X^{i},\partial_{a}X^{i}\rangle \sum_{j} (X^{j})^{2}
\end{eqnarray}
\begin{eqnarray}
\langle[X^{a},X^{b},X^{c}],[X^{a},X^{b},X^{c}]\rangle&=&
-\lambda^{2}
(F^{abc}_{\alpha\beta\gamma})(F^{abc}_{\alpha\beta\eta})\left(\langle[T^{\alpha},T^{\beta},T^{\gamma}],[T^{\alpha},T^{\beta},T^{\eta}]
\rangle\right)\nonumber \\
&=&-\lambda^{2}
(F^{abc}_{\alpha\beta\gamma})(F^{abc}_{\alpha\beta\eta})f^{\alpha
\beta \gamma}_{\sigma}h^{\sigma \kappa}f^{\alpha \beta
\eta}_{\kappa} \langle T^{\gamma},T^{\eta}\rangle \nonumber \\&=& -\lambda^{2}
(F^{abc}_{\alpha\beta\gamma})(F^{abc}_{\alpha\beta\eta})\delta^{\kappa
\sigma} \langle T^{\gamma},T^{\eta}\rangle= -\lambda^{2} \langle
F^{abc},F^{abc}\rangle \nonumber\\
\nonumber\\ F_{abc}&=&\partial_{a} A_{bc}-\partial_{b}
A_{ca}+\partial_{c}
A_{ab}. \nonumber \\
&&\nonumber \\
 i,j&=&p+1,..,10\quad a,b=0,1,...p\quad m,n=0,..,10~~ \label{s13}
\end{eqnarray}
where $A_{ab}$ is $2$-form gauge field which lives on the brane
and $\lambda=2\pi l_{s}^{2}$. Substituting  above mappings in
action (\ref{s12}), we obtain:

\begin{eqnarray}
 &&S=-\frac{T_{Mp}}{2}\int d^{p+1}x \sum_{n=1}^{p}\beta_{n} \tilde{\chi}^{\mu_{0}}_{[\mu_{0}}\tilde{\chi}^{\mu_{1}}_{\mu_{1}}
 ...\tilde{\chi}^{\mu_{n}}_{\mu_{n}]}\label{s14}
\end{eqnarray}
where
\begin{eqnarray}
 &&\tilde{\chi}^{\mu}_{\nu}\equiv
\sqrt{g^{\mu\rho}\partial_{\rho}X^{i}\partial_{\nu}X^{j}\eta_{ij}\sum_{j}
(X^{j})^{2}- \lambda^{2}\langle F^{\mu bc},F_{\nu bc}\rangle
+\langle[X^{k},X^{i},X^{j}],[X_{k},X_{i},X_{j}]\rangle}\label{s15}
\end{eqnarray}

In fact, equation (\ref{s15}) is one approximation of following
Lagrangian \cite{m7,m8,m15,m16,m17,m18}:

\begin{eqnarray}
 \tilde{\chi}= STr \sqrt{-det(P_{abc}[ E_{mnl}
+E_{mij}(Q^{-1}-\delta)^{ijk}E_{kln}]- \lambda
F_{abc})det(Q^{i}_{j,k})} \label{s16}
\end{eqnarray}
where

\begin{eqnarray}
 E_{mnl}^{\alpha,\beta,\gamma} &=& G_{mnl}^{\alpha,\beta,\gamma} + B_{mnl}^{\alpha,\beta,\gamma}, \nonumber\\
 Q^{i}_{j,k} &=& \delta^{i}_{j,k} + i\lambda[X^{j}_{\alpha}T^{\alpha},X^{k}_{\beta}T^{\beta},X^{k'}_{\gamma}T^{\gamma}]
 E_{k'jl}^{\alpha,\beta,\gamma},\nonumber\\F_{abc}&=&\partial_{a} A_{bc}-\partial_{b} A_{ca}+\partial_{c}
A_{ab}. \label{s17}
\end{eqnarray}
and $X^{M}=X^{M}_{\alpha}T^{\alpha}$, $A_{ab}$ is $2$-form gauge
field,

\begin{eqnarray}
 &&[T^{\alpha}, T^{\beta}, T^{\gamma}]=
 f^{\alpha \beta \gamma}_{\eta}T^{\eta} \nonumber \\&& [X^{M},X^{N},X^{L}]=
 [X^{M}_{\alpha}T^{\alpha},X^{N}_{\beta}T^{\beta},X^{L}_{\gamma}T^{\gamma}]
\label{s18}
\end{eqnarray}
where $\lambda=2\pi l_{s}^{2}$,
$G_{mnl}=g_{mn}\delta^{n'}_{n,l}+\partial_{m}X^{i}\partial_{n'}X^{i}\sum_{j}
(X^{j})^{2}\delta^{n'}_{n,l} $ and $X^{i}$ refers to scalar field.
Here $a,b=0,1,...,p$ denote the world-volume indices of the
$Mp$-branes, $i,j,k = p+1,...,9$ refer to indices of the
transverse space, and $m$, $n$ denote the eleven-dimensional
space-time indices.
It is clear from above calculations that by linking M0-branes and formation an Mp-brane,  two form gauge fields are produced that
may have the role of tensor mode of graviton in real world. Also,
scalar fields  may play the role of scalar mode of graviton. We
can write:

\begin{eqnarray}
&&A^{ab}=g^{ab}=h^{ab} ~ ~  and  ~ ~ a,b,c=\mu,\nu,\lambda
\Rightarrow \nonumber\\&& F_{abc}=\partial_{a} A_{bc}-\partial_{b}
A_{ca}+\partial_{c}
A_{ab}=2(\partial_{\mu}g_{\nu\lambda}+\partial_{\nu}g_{\mu\lambda}-\partial_{\lambda}g_{\mu\nu})=
2\Gamma_{\mu\nu\lambda}\nonumber\\&&\nonumber\\&&\langle
F^{\rho}\smallskip_{\sigma\lambda},F^{\lambda}\smallskip_{\mu\nu}\rangle=
\langle[X^{\rho},X_{\sigma},X_{\lambda}],[X^{\lambda},X_{\mu},X_{\nu}]\rangle=\nonumber\\&&
[X_{\nu},[X^{\rho},X_{\sigma},X_{\mu}]]-[X_{\mu},[X^{\rho},X_{\sigma},X_{\nu}]]+[X^{\rho},X_{\lambda},X_{\nu}]
[X^{\lambda},X_{\sigma},X_{\mu}]
-[X^{\rho},X_{\lambda},X_{\mu}][X^{\lambda},X_{\sigma},X_{\nu}]=\nonumber\\&&\partial_{\nu}
\Gamma^{\rho}_{\sigma\mu}-\partial_{\mu}\Gamma^{\rho}_{\sigma\nu}+\Gamma^{\rho}_{\lambda\nu}
\Gamma^{\lambda}_{\sigma\mu}-\Gamma^{\rho}_{\lambda\mu}\Gamma^{\lambda}_{\sigma\nu}
=R^{\rho}_{\sigma\mu\nu}\label{s19}
\end{eqnarray}
and
\begin{eqnarray}
&&\kappa^{\mu}_{\nu}=\delta^{\mu}_{\nu}-\sqrt{\delta^{\mu}_{\nu}-H^{\mu}_{\nu}}\nonumber\\
&&H_{\mu\nu}=
g_{\mu\nu}-\eta_{mn}\partial_{\mu}X^{m}\partial_{\nu}X^{n}\nonumber\\
&&H_{\mu\nu}=h_{\mu\nu}+2\Pi_{\mu\nu}-\eta^{\alpha\beta}\Pi_{\mu\alpha}\Pi_{\beta\nu}\nonumber\\
&&X^{m}=x^{m}-\eta^{m\mu}\partial_{\mu}\pi \nonumber\\
&&\Pi_{\mu\nu}=\partial_{\mu}\partial_{\nu}\pi \label{s20}
\end{eqnarray}
Here $\pi$ denotes the scalar mode  and $h^{ab}$ refers to the
tensor mode of graviton. Until now, many papers have discussed
that there can be existed  antisymmetric metric \cite{fd1,fd2}and
antisymmetric gravity \cite{fd3,fd4}.  On the other hand, when we
speak of BIon, our metric is consisted of metrics of two universe
branes which are connected by a wormhole. One can construct
antisymmetric form of metric by combining these three metrics.
Thus, we can put the antisymmetric metric equal to two form gauge
fields in M-theory. Obviously, non-commutative bracket for two
form fields produces the exact form of curvature tensor. Also, by
linking scalars to branes, they play the role of scalar mode of
graviton. In previous considerations, it has been found that there
is a relation between $\kappa$ and curvature scalars ($R$)
\cite{m9}:

\begin{eqnarray}
&&\delta^{\rho\sigma}_{\mu\nu}\kappa^{\mu}_{\rho}\kappa^{\nu}_{\sigma}=R
\label{s21}
\end{eqnarray}

Thus, curvature scalars and tensors in gravity  can be obtained
from the non-commutative brackets in M-theory. At this stage, we
can derive the form of $\tilde{\chi}$ in equation (\ref{s16}) in
terms of curvature scalars and tensors in gravity. We have:

\begin{eqnarray}
&&
\det(Z)=\delta^{a_{1},a_{2}...a_{n}}_{b_{1}b_{2}....b_{n}}Z^{b_{1}}_{a_{1}}...Z^{b_{n}}_{a_{n}}
\quad a,b,c=\mu,\nu,\lambda \nonumber\\\nonumber\\
&&Z_{abc}=P_{abc}[ E_{mnl} +E_{mij}(Q^{-1}-\delta)^{ijk}E_{kln}]+
\lambda F_{abc}\nonumber\\&&\nonumber\\
&&\det(Z)=\det\left(P_{abc}[ E_{mnl}
+E_{mij}(Q^{-1}-\delta)^{ijk}E_{kln}]\right)+\lambda^{2}\det(F)\label{s22}\end{eqnarray}

Using the above relation, we can calculate the relevant terms of
determinant in action (\ref{s16}) separately. Applying relations
(\ref{s19},\ref{s20},\ref{s21}) in determinants (\ref{s22}), we
obtain:

\begin{eqnarray}
\det(F)=\delta_{\rho\sigma}^{\mu\nu}\langle
F^{\rho\sigma}\smallskip_{\lambda},F^{\lambda}\smallskip_{\mu\nu}\rangle
=\delta_{\rho\sigma}^{\mu\nu}R^{\rho\sigma}_{\mu\nu}\label{s23}\end{eqnarray}

\begin{eqnarray}
\det(P_{abc}[ E_{mnl}
+E_{mij}(Q^{-1}-\delta)^{ijk}E_{kln}])&=&\delta_{\rho\sigma}^{\mu\nu}
\Biggl[(g^{\mu}_{\rho}g^{\nu}_{\sigma}+ g^{\nu}_{\sigma}\langle
\partial^{\mu}X^{i},\partial_{\rho}X^{j}\rangle \sum (X^{i})^{2}+..)
\nonumber\\&&-\frac{(g^{\mu}_{\rho}g^{\nu}_{\sigma}+
g^{\nu}_{\sigma}\langle
\partial^{\mu}X^{i},\partial_{\rho}X^{j}\rangle \sum (X^{i})^{2}+...)}{[(\lambda)^{2}
\det([X^{j}_{\alpha}T^{\alpha},X^{k}_{\beta}T^{\beta},X^{k'}_{\gamma}T^{\gamma}])]}\Biggr]
\nonumber\\&=&\delta_{\rho\sigma}^{\mu\nu}[\kappa_{\mu}^{\rho}\kappa_{\nu}^{\sigma}
\sum (X^{i})^{2}]
\left(1-\frac{1}{\left[(\lambda)^{2}\det([X^{j}_{\alpha}T^{\alpha},X^{k}_{\beta}T^{\beta},X^{k'}_{\gamma}T^{\gamma}])\right]}\right)
\nonumber\\&=&\delta_{\rho\sigma}^{\mu\nu}\left[\kappa_{\mu}^{\rho}\kappa_{\nu}^{\sigma}
\sum
(X^{i})^{2}\right]\left(1-\frac{1}{m_{g}^{2}}\right)\label{s24}
\end{eqnarray}
where $m_{g}^{2}= [(\lambda)^{2}\det([X^{j}_{\alpha}T^{\alpha},X^{k}_{\beta}T^{\beta},X^{k'}_{\gamma}T^{\gamma}])]
$ denotes the square of graviton mass. It is clear that the
graviton mass depends on the scalars which interact with branes. Since, by colliding scalars with branes, they become graviton. Using this definition, we can derive another term of the determinant:

\begin{eqnarray}
\det(Q)&\sim&
(i)^{2}(\lambda)^{2}\det([X^{j}_{\alpha}T^{\alpha},X^{k}_{\beta}T^{\beta},X^{k'}_{\gamma}T^{\gamma}])\det(E)
\nonumber\\&\sim&
-[(\lambda)^{2}\det([X^{j}_{\alpha}T^{\alpha},X^{k}_{\beta}T^{\beta},X^{k'}_{\gamma}T^{\gamma}])]\det(g)=m_{g}^{2}\det(g)
\label{s25}
\end{eqnarray}
Applying relations (\ref{s19}) ,(\ref{s20}), (\ref{s21}),
(\ref{s22}), (\ref{s23}) ,(\ref{s24}), and (\ref{s25}) in the
action (\ref{s16}) and defining $\sum (X^{i})^{2}\rightarrow
F(X)$, we get:

\begin{eqnarray}
 \tilde{\chi}&=&
\Bigl[\sqrt{-g}\Big(\delta^{\rho\sigma}_{\mu\nu}[\kappa_{\mu}^{\rho}\kappa_{\nu}^{\sigma}
\sum (X^{i})^{2}]-
m_{g}^{2}\delta^{\rho\sigma}_{\mu\nu}\left(R_{\rho\sigma}^{\mu\nu}+[\kappa_{\mu}^{\rho}\kappa_{\nu}^{\sigma}
\sum (X^{i})^{2}]\right)\Big)\Bigr]\nonumber\\&=& -
\Bigl[\sqrt{-g}\Big(F(X)(1-m_{g}^{2})R-
m_{g}^{2}\lambda^{2}\delta^{\rho\sigma}_{\mu\nu}R_{\rho\sigma}^{\mu\nu})\Big)\Bigr]
\label{s26}
\end{eqnarray}
Substituting equation (\ref{s26}) in equation (\ref{s14}) yields:
\begin{eqnarray}
S_{Mp} &=& -\frac{T_{Mp}}{2}\int d^{p+1}x \sum_{n=1}^{p}\beta_{n}
\tilde{\chi}^{\mu_{0}}_{[\mu_{0}}\tilde{\chi}^{\mu_{1}}_{\mu_{1}}
 ...\tilde{\chi}^{\mu_{n}}_{\mu_{n}]}=\nonumber\\&&
-\frac{T_{Mp}}{2} \int dt \int d^{p}\sigma \sum_{n=1}^{p}
\beta_{n}
\delta^{a_{1},a_{2}...a_{n}}_{b_{1}b_{2}....b_{n}}\nonumber\\
&& \Bigl[\sqrt{-g_{1}}\Bigl(F(X)(1-m_{g}^{2})R-
m_{g}^{2}\lambda^{2}\delta^{\rho_{1}\sigma_{1}}_{\mu_{1}\nu_{1}}R_{\rho_{1}\sigma_{1}}^{\mu_{1}\nu_{1}}\Bigr)\Bigr]
^{b_{1}}_{a_{1}}\times
\nonumber\\&&.......\times\nonumber\\
&& \Biggl[\sqrt{-g_{p}}\Bigl(F(X)(1-m_{g}^{2})R-
m_{g}^{2}\lambda^{2}\delta^{\mu_{p}\nu_{p}}_{\rho_{p}\sigma_{p}}R_{\rho_{p}\sigma_{p}}^{\mu_{p}\nu_{p}}\Bigr)\Bigr]
^{b_{p}}_{a_{p}}\nonumber\\&=&-(T_{Mp})\int dt \int d^{p}\sigma
\Biggl[\sqrt{-g}\Bigl(\sum_{n=1}^{p}\beta_{n}F(X)^{n}(1-m_{g}^{2})^{n}
R^{n}-\nonumber\\&&
\sum_{n=1}^{p}m_{g}^{2n}\lambda^{2n}\delta^{\rho_{1}\sigma_{1}...\rho_{n}\sigma_{n}}_{\mu_{1}\nu_{1}...\mu_{n}\nu_{n}}
R_{\rho_{1}\sigma_{1}}^{\mu_{1}\nu_{1}}...R_{\rho_{n}\sigma_{n}}^{\mu_{n}\nu_{n}}
\Bigr)\Biggr] \label{s27}
\end{eqnarray}

This equation shows that  all terms in nonlinear gravity theories
like  Lovelock gravity\cite{m2} and F(R)-gravity \cite{m21} can be
extracted from actions in M-theory. Also, by increasing dimensions
of Mp-brane, higher orders of gauge fields and scalars are created
which appear as higher orders of curvature tensors and scalars in
the effective potential between branes and anti-branes. These
nonlinear gravity terms cause production of absorption force
between branes and generation of confinement.

Till now, we have shown that equation (\ref{s14}) for branes and
equation (\ref{s27}) for Lovelock gravity are the same. Now, we
obtain the gravitational energy in this system. In our model,
universes are placed on the M3-brane and interact with anti-branes
which live on anti-M3-branes via M2-branes. Also, in this model,
gauge fields live on the brane and scalars live in extra
dimensions and by increasing the length of brane, gauge fields
grow and scalars decrease. Using the equation (\ref{s12})and
assuming the length of M2 between two M3 be $l_{2}$ and the length
of each M3 be $l_{3}$, we can derive the relevant action for the
system of M3-M2:

\begin{eqnarray}
&&  A^{ab}\rightarrow l_{3} \quad X^{i}\rightarrow l_{2} \quad
\text{for M3-brane}
\nonumber\\&& A^{ab}\rightarrow l_{2} \quad X^{i}\rightarrow l_{5} \quad \text{for M2-brane}\nonumber\\
&&  S_{tot}=S_{M3}+S_{M2}+S_{anti-M3}=\nonumber\\&&-2T_{M3} \int
d^{4}\sigma \Sigma_{n=1}^{3}
\beta_{n}(l_{2}^{6}+l_{2}'^{2}l_{2}^{2}-\lambda^{2}l_{3}'^{2})^{\frac{n}{2}}\nonumber\\&&-2T_{M2}
\int d^{3}\sigma \Sigma_{n=1}^{2}\beta_{n}
(l_{3}^{6}+l_{3}'^{2}l_{3}^{2}-\lambda^{2}l_{2}'^{2})^{\frac{n}{2}}\simeq\nonumber\\&&-2T_{M3}
\int dt l_{3}^{3} \Sigma_{n=1}^{3}
\beta_{n}(l_{2}^{6}+l_{2}'^{2}l_{2}^{2}-\lambda^{2}l_{3}'^{2})^{\frac{n}{2}}\nonumber\\&&-2T_{M2}
\int dt l_{2}^{2} \Sigma_{n=1}^{2}\beta_{n}
(l_{3}^{6}+l_{3}'^{2}l_{3}^{2}-\lambda^{2}l_{2}'^{2})^{\frac{n}{2}}\label{s28}
\end{eqnarray}

The equation of motion for this equation is:

\begin{eqnarray}
&&-2T_{M3} \Sigma_{n=1}^{3} \beta_{n}[ l_{3}^{3}(
l_{2}'l_{2}^{2})(l_{2}^{6}+l_{2}'^{2}l_{2}^{2}-\lambda^{2}l_{3}'^{2})^{\frac{n}{2}-1}]'\nonumber\\&&
+2T_{M2} \Sigma_{n=1}^{2} \beta_{n}[(\lambda^{2} l_{2}^{2}
l_{2}'))(l_{3}^{6}+l_{3}'^{2}l_{3}^{2}-\lambda^{2}l_{2}'^{2})^{\frac{n}{2}-1}]'\nonumber\\&&
+2T_{M3} \Sigma_{n=1}^{3} \beta_{n}[ l_{3}^{3}(
l_{2}'^{2}l_{2}+3l_{2}^{3})(l_{2}^{6}+l_{2}'^{2}l_{2}^{2}-\lambda^{2}l_{3}'^{2})^{\frac{n}{2}-1}]\nonumber\\&&
-2T_{M2} \Sigma_{n=1}^{2} \beta_{n}[(
l_{2}))(l_{3}^{6}+l_{3}'^{2}l_{5}^{2}-\lambda^{2}l_{2}'^{2})^{\frac{n}{2}}]=0
\label{s29}
\end{eqnarray}

\begin{eqnarray}
&&-2T_{M2} \Sigma_{n=1}^{2} \beta_{n}[ l_{2}^{2}(
l_{3}'l_{3}^{2})(l_{3}^{6}+l_{3}'^{2}l_{3}^{2}-\lambda^{2}l_{2}'^{2})^{\frac{n}{2}-1}]'\nonumber\\&&
+2T_{M3} \Sigma_{n=1}^{3} \beta_{n}[(\lambda^{2} l_{3}^{3}
l_{3}'))(l_{2}^{6}+l_{2}'^{2}l_{2}^{2}-\lambda^{2}l_{3}'^{2})^{\frac{n}{2}-1}]'\nonumber\\&&
+2T_{M2} \Sigma_{n=1}^{2} \beta_{n}[ l_{2}^{2}(
l_{3}'^{2}l_{3}+3l_{3}^{3})(l_{3}^{6}+l_{3}'^{2}l_{3}^{2}-\lambda^{2}l_{2}'^{2})^{\frac{n}{2}-1}]\nonumber\\&&
-2T_{M3} \Sigma_{n=1}^{3} \beta_{n}[(3
l_{3}^{2}))(l_{2}^{6}+l_{2}'^{2}l_{2}^{2}-\lambda^{2}l_{3}'^{2})^{\frac{n}{2}}]=0
\label{ss29}
\end{eqnarray}

The approximate solutions of these equations are:

\begin{eqnarray}
&&l_{3} \simeq \sum_{n=1}^{3}\beta_{n}\frac{1}{\lambda^{n}
T_{M3}^{6n}}[ln(1-\frac{t}{t_{s}})]^{\frac{n}{2}}e^{n\lambda\frac{t}{t_{s}}}\nonumber\\&&
l_{2} \simeq \sum_{n=1}^{2}\beta_{n}\frac{1}{\lambda^{n}
T_{M2}^{3n}}(1-\frac{t}{t_{s}})^{\frac{n}{2}}e^{-n\lambda\frac{t}{t_{s}}}
\label{s30}
\end{eqnarray}
where $t_{s}$ is time of collision between branes. These results
show that by passing time, the length of M2 between two M3-branes
decreases and shrinks to zero, while the length of M3 increases to
large values. This means that M2 dissolves in M3-branes and leads
to their expansion. In our model, universes are placed on the
M3-branes and connect with each other via M2-branes. By passing
time, M2 disappears and nonlinear gravity emerges which is the
main cause of confinement between branes and anti-branes. In fact,
M2 plays the role of wormhole as indicated in \cite{m1}. Now,
using equations (\ref{s28} and \ref{s30}), we  obtain the
effective potential between branes and anti-branes:

\begin{eqnarray}
&&t\rightarrow t_{s}\Rightarrow l_{3}\simeq l_{3}'  \simeq
\frac{1}{l_{2}} \quad l_{2}' \simeq \frac{1}{l_{2}}\nonumber\\&&
V_{brane-antibrane}=2T_{M3}  l_{2}^{-3} \Sigma_{n=1}^{3}
\beta_{n}(l_{2}^{6}+1-\lambda^{2}l_{2}^{-2})^{\frac{n}{2}}\nonumber\\&&+2T_{M2}
l_{2}^{2} \Sigma_{n=1}^{2}\beta_{n}
(l_{2}^{-6}+l_{2}^{-4}-\lambda^{2}l_{2}^{-2})^{\frac{n}{2}}\simeq
\nonumber\\&& c_{1}l_{2}-c_{2}l_{2}^{-1}-c_{3}l_{2}^{-3}+...
\label{s31}
\end{eqnarray}

where $c_{i}$'s are some coefficients which  depend on the tension
of branes and $\lambda$. This potential is very similar to
prediction of QCD for potential between quarks and anti-quarks
\cite{m22}. This means that usual potential between branes which
is produced by dissolving M2-branes in M3-branes can be used in
QCD. On the other hand, we have shown that by closing branes and
disappearing M2, Lovelock gravity emerges. Thus, the Lovelock
gravity may be the main cause of potential in brane systems.  Now,
we want to consider the evolution of gravitational energy in BIon.
First, we show that by closing branes to each other, BIon becomes
thermal and transits to black hole. To this end, we begin with
equation of motion of fields between branes and use of the method
in \cite{m3}:

\begin{eqnarray}
&& -\frac{\partial^{2} A}{\partial t^{2}} + \frac{\partial^{2}
A}{\partial z^{2}}=0 \label{t10}
\end{eqnarray}

where z is the transverse dimension between two M3-branes. By
using the following re-parameterizations

\begin{eqnarray}
&& \rho = \frac{z^{2}}{w} ,  \nonumber \\&& w=
\frac{V_{brane-antibrane}}{2E_{system}}\nonumber \\&& \bar{\tau} =
\gamma\int_{0}^{t} d\tau' \frac{w}{\dot{w}} - \gamma
\frac{z^{2}}{2}\label{t11}
\end{eqnarray}

and doing the  following calculations:

\begin{eqnarray}
\left [\left\{(\frac{\partial \bar{\tau}}{\partial t})^{2} -
(\frac{\partial \bar{\tau}}{\partial
z})^{2}\right\}\frac{\partial^{2}}{\partial
t^{2}}+\left\{(\frac{\partial \rho}{\partial z})^{2} -
(\frac{\partial \rho}{\partial t})^{2} \right
\}\frac{\partial^{2}}{\partial \rho^{2}}\right]A=0\label{t12}
\end{eqnarray}

we get \cite{m3}:

\begin{eqnarray}
&& (-g)^{-1/2}\frac{\partial}{\partial
x_{\mu}}[(-g)^{1/2}g^{\mu\nu}]\frac{\partial}{\partial
x_{\upsilon}}A=0\label{t13}
\end{eqnarray}

where $x_{0}=\bar{\tau}$, $x_{1}=\rho$ and the metric elements are
obtained as:

\begin{eqnarray}
&& g^{\bar{\tau}\bar{\tau}}\sim
-\frac{1}{\beta^{2}}(\frac{w'}{w})^{2}\frac{(1-(\frac{w}{w'})^{2}\frac{1}{z^{4}})}{(1+(\frac{w}{w'})^{2}\frac{(1+\gamma^{-2})}{z^{4}})^{1/2}}\nonumber
\\&&g^{\rho\rho}\sim -(g^{\bar{\tau}\bar{\tau}})^{-1}\label{t14}
\end{eqnarray}

At this stage, we  compare these elements with the line elements
of a thermal BIon \cite{m3,m23}:

\begin{eqnarray}
&& ds^{2}= D^{-1/2}\bar{H}^{-1/2}(-f
dt^{2}+dx_{1}^{2})+D^{1/2}\bar{H}^{-1/2}(dx_{2}^{2}+dx_{3}^{2})+D^{-1/2}\bar{H}^{1/2}(f^{-1}
dr^{2}+r^{2}d\Omega_{5})\nonumber
\\&&\label{t15}
\end{eqnarray}

where

\begin{eqnarray}
&&f=1-\frac{r_{0}^{4}}{r^{4}},\nonumber
\\&&\bar{H}=1+\frac{r_{0}^{4}}{r^{4}}sinh^{2}\alpha \nonumber
\\&&D^{-1}=cos^{2}\varepsilon + H^{-1}sin^{2}\varepsilon \nonumber
\\&&cos\varepsilon =\frac{1}{\sqrt{1+\frac{\beta^{2}}{l_{2}^{4}}}}
\label{t16}
\end{eqnarray}

Comparing the line elements of (\ref{t16}) equal to line elements
of (\ref{t14}), we obtain \cite{m3,m23}:

\begin{eqnarray}
&&f=1-\frac{r_{0}^{4}}{r^{4}}\sim
1-\left(\frac{w}{w'}\right)^{2}\frac{1}{z^{4}},\nonumber
\\&&\bar{H}=1+\frac{r_{0}^{4}}{r^{4}}sinh^{2}\alpha \sim 1+\left(\frac{w}{w'}\right)^{2}\frac{(1+\gamma^{-2})}{z^{4}} \nonumber
\\&&D^{-1}=cos^{2}\varepsilon + \bar{H}^{-1}sin^{2}\varepsilon\simeq1\nonumber
\\&&
\Rightarrow r\sim z,r_{0}\sim
\left(\frac{w}{w'}\right)^{1/2},(1+\gamma^{-2})\sim sinh^{2}\alpha
\nonumber
\\&&cosh^{2}\alpha = \frac{3}{2}\frac{cos\frac{\delta}{3} +
\sqrt{3}sin\frac{\delta}{3}}{cos\delta}\nonumber
\\&& cos\delta \equiv \overline{T}^{4}\sqrt{1 +
\frac{k^{2}}{l_{2}^{4}}},\, \overline{T} \equiv
\frac{T}{T_{c}}\label{t17}
\end{eqnarray}

The temperature of BIon is ${\displaystyle T=\frac{1}{\pi r_{0}
cosh\alpha}}$, see \cite{m3,m23} for details. Consequently, the
temperature of a BIon can be derived as:

\begin{eqnarray}
&& T=\frac{1}{\pi r_{0}
cosh\alpha}=\frac{\gamma}{\pi}(\frac{w'}{w})^{1/2}\sim
\frac{\gamma}{\pi}(\frac{V'}{V})^{1/2}\sim \nonumber
\\&&\frac{\gamma}{\pi}\left(\frac{c_{1}-c_{2}l_{2}^{-2}-c_{3}l_{2}^{-4}}{c_{1}l_{2}-c_{2}l_{2}^{-1}-c_{3}l_{2}^{-3}}\right)^{1/2} \label{t18}
\end{eqnarray}

Because $\gamma$ depends on the temperature, we can write:

\begin{eqnarray}
&&\gamma = \frac{1}{cosh\alpha} \sim
\frac{2cos\delta}{3\sqrt{3}-cos\delta
-\frac{\sqrt{3}}{6}cos^{2}\delta}\sim \nonumber
\\&&\frac{2\bar{T}^{4}\sqrt{1+\frac{k^{2}}{l_{2}^{4}}}}{3\sqrt{3}-\bar{T}^{4}\sqrt{1+\frac{k^{2}}{l_{2}^{4}}}
-\frac{\sqrt{3}}{6}\bar{T}^{8}(1+\frac{k^{2}}{l_{2}^{4}})}
\label{t19}
\end{eqnarray}

Using Eqs. (\ref{t18}) and (\ref{t19}) , we can approximate the
explicit form of  the size of M2 in terms of the temperature:

\begin{eqnarray}
&& l_{2}\simeq
T^{-\frac{2}{3}}[\frac{6}{k\sqrt{3}}-\bar{T}^{2}]^{\frac{2}{3}}
\label{t20}
\end{eqnarray}

It is clear that temperature of system at colliding point has the
following relation with critical temperature $T_{c}$:

\begin{eqnarray}
&& T_{\text{colliding
point}}=\sqrt{\frac{6}{k\sqrt{3}}}T_{c}\label{t21}
\end{eqnarray}

By definition of $k=\frac{\sqrt{3}}{6}$, temperature of colliding
point and critical temperature become equal ($T_{\text{colliding
point}}=T_{c}$). Now, we can obtain the Brown-York energy during
the expansion branch. To this end, we use of the method in
\cite{s5}. The metric (\ref{t15})is assumed to be asymptotically
flat, i.e., in the limit $r,z\rightarrow\infty$ and
$l_{2}\rightarrow\infty$, parameters of BIon go to unity ( $
D,H,f\rightarrow\infty $) giving the flat Minkowskian metric. To
match with the method in \cite{s5}, we redefine the line elements
in (\ref{t15}) as

\begin{eqnarray}
&& ds_{BIon}^{2}=ds_{brane}^{2}+ds_{BH}^{2}\nonumber
\\&&ds_{BH}^{2}=- D^{-1/2}\bar{H}^{-1/2}f
dt^{2}+D^{-1/2}\bar{H}^{1/2}f^{-1}
dr^{2}+r^{2}d\Omega_{2}\nonumber
\\&&ds_{brane}^{2}=
D^{-1/2}\bar{H}^{-1/2}dx_{1}^{2}+D^{1/2}\bar{H}^{-1/2}(dx_{2}^{2}+dx_{3}^{2})+D^{-1/2}\bar{H}^{1/2}r^{2}d\Omega_{3}\label{tt15}
\end{eqnarray}

 where have used of this fact that BIon contains two branes which are connected by a black hole. We focus on the line element of black hole ($ds_{BH}^{2}$).
 The hypersurface $\Sigma$ is taken to be a t = constant
hypersurface and B as a r = constant surface. The two-surface $B$
is a r = constant hypersurface within $\Sigma$. Then, the unit
timelike normal $u_{a}$ to $\Sigma$ and unit spacelike normal
$n_{a}$ to $B$ turn out to be

\begin{eqnarray}
&&u_{a}=-\sqrt{D^{-1/2}\bar{H}^{-1/2}f}(1,0,0,0,0,0,0,0,0,0,0);\quad
u^{a}=\sqrt{D^{1/2}\bar{H}^{-1/2}f^{-1}}(1,0,0,0,0,0,0,0,0,0,0)\nonumber\\&&\eta_{a}=\sqrt{D^{-1/2}\bar{H}^{1/2}f^{-1}}(0,1,0,0,0,0,0,0,0,0,0);\quad
\eta^{a}=\sqrt{D^{1/2}\bar{H}^{-1/2}f}(0,1,0,0,0,0,0,0,0,0,0)\label{Q1}
\end{eqnarray}

Having derived the normal to $\Sigma$ and B we can now obtain the
corresponding extrinsic curvatures, and in particular for the
latter we get,

\begin{eqnarray}
&&
k=-\frac{1}{\sqrt{h}}\partial_{\mu}(\sqrt{h}\eta^{\mu})=-\partial_{r}\eta^{r}-\eta^{r}\partial_{r}ln\sqrt{h}=\nonumber\\&&
-\partial_{r}(\sqrt{D^{1/2}\bar{H}^{-1/2}f})-\sqrt{D^{1/2}\bar{H}^{-1/2}f}
ln(\frac{r^{2}sin\theta}{\sqrt{D^{1/2}\bar{H}^{-1/2}f}})=-\frac{2\sqrt{D^{1/2}\bar{H}^{-1/2}f}}{r}\label{Q2}
\end{eqnarray}

The embedding of $B$ in a flat spacetime is trivial and the trace
of extrinsic curvature $k_{0}$ is simply $-\frac{2}{r}$.  Then
using the curvature in Eq. (\ref{Q2})and the Brown-York energy
defined in Eq. (\ref{p1}), we obtain:

\begin{eqnarray}
&&E_{BY}=\frac{1}{4\pi} \int d\theta d\phi r^{2}sin\theta
\frac{1}{r}(1-\sqrt{D^{1/2}\bar{H}^{-1/2}f})=r(1-\sqrt{D^{1/2}\bar{H}^{-1/2}f})=\nonumber\\&&r(1-\sqrt{[cos^{2}\varepsilon
+
[1+\frac{r_{0}^{4}}{r^{4}}sinh^{2}\alpha]^{-1}sin^{2}\varepsilon]^{-1/2}[1+\frac{r_{0}^{4}}{r^{4}}sinh^{2}\alpha]^{-1/2}[1-\frac{r_{0}^{4}}{r^{4}}]})
\label{Q3}
\end{eqnarray}

Substituting equations
(\ref{s30},\ref{t16},\ref{t17},\ref{t18},\ref{t19}and \ref{t20})in
equation (\ref{Q3})  in the limit $r\sim z\sim l_{2}\rightarrow
0$, we obtain:

\begin{eqnarray}
&&E_{BY}=l_{2}+\Big(\frac{l_{2,0}^{4}}{l_{2}^{3}}
[\frac{1}{1+\frac{\beta^{2}}{l_{2}^{4}}} +
[1+\nonumber\\&&\frac{l_{2,0}^{4}}{l_{2}^{4}}[1-(\frac{3\sqrt{3}-\bar{T}^{4}\sqrt{1+\frac{k^{2}}{l_{2}^{4}}}
-\frac{\sqrt{3}}{6}\bar{T}^{8}(1+\frac{k^{2}}{l_{2}^{4}})}{2\bar{T}^{4}\sqrt{1+\frac{k^{2}}{l_{2}^{4}}}})^{2}]]^{-1}[1-\frac{1}{1+\frac{\beta^{2}}{l_{2}^{4}}}]]^{-1/2}
\times
\nonumber\\&&[1+\frac{l_{2,0}^{4}}{l_{2}^{4}}[(\frac{3\sqrt{3}-\bar{T}^{4}\sqrt{1+\frac{k^{2}}{l_{2}^{4}}}
-\frac{\sqrt{3}}{6}\bar{T}^{8}(1+\frac{k^{2}}{l_{2}^{4}})}{2\bar{T}^{4}\sqrt{1+\frac{k^{2}}{l_{2}^{4}}}})^{2}]]^{-1/2}\Big)^{\frac{1}{2}}\sim
\nonumber\\&&[\sum_{n=1}^{2}\beta_{n}\frac{1}{\lambda^{n}
T_{M2}^{3n}}(1-\frac{t}{t_{s}})^{\frac{n}{2}}e^{-n\lambda\frac{t}{t_{s}}}
]+\Big(\frac{l_{2,0}^{4}}{[\sum_{n=1}^{2}\beta_{n}\frac{1}{\lambda^{n}
T_{M2}^{3n}}(1-\frac{t}{t_{s}})^{\frac{n}{2}}e^{-n\lambda\frac{t}{t_{s}}}
]^{3}}\times\nonumber\\&&
[\frac{1}{1+\frac{\beta^{2}}{[\sum_{n=1}^{2}\beta_{n}\frac{1}{\lambda^{n}
T_{M2}^{3n}}(1-\frac{t}{t_{s}})^{\frac{n}{2}}e^{-n\lambda\frac{t}{t_{s}}}
]^{4}}} +
[1+\frac{l_{2,0}^{4}}{[\sum_{n=1}^{2}\beta_{n}\frac{1}{\lambda^{n}
T_{M2}^{3n}}(1-\frac{t}{t_{s}})^{\frac{n}{2}}e^{-n\lambda\frac{t}{t_{s}}}
]^{4}}[1-\nonumber\\&&(\frac{3\sqrt{3}-\bar{T}^{4}\sqrt{1+\frac{k^{2}}{[\sum_{n=1}^{2}\beta_{n}\frac{1}{\lambda^{n}
T_{M2}^{3n}}(1-\frac{t}{t_{s}})^{\frac{n}{2}}e^{-n\lambda\frac{t}{t_{s}}}
]^{4}}}
-\frac{\sqrt{3}}{6}\bar{T}^{8}(1+\frac{k^{2}}{[\sum_{n=1}^{2}\beta_{n}\frac{1}{\lambda^{n}
T_{M2}^{3n}}(1-\frac{t}{t_{s}})^{\frac{n}{2}}e^{-n\lambda\frac{t}{t_{s}}}
]^{4}})}{2\bar{T}^{4}\sqrt{1+\frac{k^{2}}{[\sum_{n=1}^{2}\beta_{n}\frac{1}{\lambda^{n}
T_{M2}^{3n}}(1-\frac{t}{t_{s}})^{\frac{n}{2}}e^{-n\lambda\frac{t}{t_{s}}}
]^{4}}}})^{2}]]^{-1}\times
\nonumber\\&&[1-\frac{1}{1+\frac{\beta^{2}}{[\sum_{n=1}^{2}\beta_{n}\frac{1}{\lambda^{n}
T_{M2}^{3n}}(1-\frac{t}{t_{s}})^{\frac{n}{2}}e^{-n\lambda\frac{t}{t_{s}}}
]^{4}}}]]^{-1/2}
[1+\frac{l_{2,0}^{4}}{[\sum_{n=1}^{2}\beta_{n}\frac{1}{\lambda^{n}
T_{M2}^{3n}}(1-\frac{t}{t_{s}})^{\frac{n}{2}}e^{-n\lambda\frac{t}{t_{s}}}
]^{4}}\times\nonumber\\&&[(\frac{3\sqrt{3}-\bar{T}^{4}\sqrt{1+\frac{k^{2}}{[\sum_{n=1}^{2}\beta_{n}\frac{1}{\lambda^{n}
T_{M2}^{3n}}(1-\frac{t}{t_{s}})^{\frac{n}{2}}e^{-n\lambda\frac{t}{t_{s}}}
]^{4}}}
-\frac{\sqrt{3}}{6}\bar{T}^{8}(1+\frac{k^{2}}{[\sum_{n=1}^{2}\beta_{n}\frac{1}{\lambda^{n}
T_{M2}^{3n}}(1-\frac{t}{t_{s}})^{\frac{n}{2}}e^{-n\lambda\frac{t}{t_{s}}}
]^{4}})}{2\bar{T}^{4}\sqrt{1+\frac{k^{2}}{[\sum_{n=1}^{2}\beta_{n}\frac{1}{\lambda^{n}
T_{M2}^{3n}}(1-\frac{t}{t_{s}})^{\frac{n}{2}}e^{-n\lambda\frac{t}{t_{s}}}
]^{4}}}})^{2}]]^{-1/2}\Big)^{\frac{1}{2}}\label{Q4}
\end{eqnarray}

This equation shows that by passing time and decreasing the
separation distance between two branes ($l_{2}$), the Brown-York
energy increases and tends to infinity at ($t=t_{s}$) (See Figure
1(Left).). This is because that by approaching M3-branes, M2
dissolves in them, nonlinear gravities like Lovelock gravity grow
and temperature and also the gravitational energy of system
increase.

At this stage, we can generalized his energy to all orders of
Lovelock gravity. Previously, it has been asserted that the
Brown-York energy for mth order Lovelock static black hole reads
as \cite{s5}:

\begin{eqnarray}
&&\overline{E}_{BY}=r^{D-2m-1}[D_{0}(1-\sqrt{\bar{f}})^{2m-1}+...+D_{s}(1-\sqrt{\bar{f}})^{2m-2s-1}(1-\bar{f})^{s}+...D_{m-1}(1-\bar{f})^{m-1}(1-\sqrt{\bar{f}})]
\label{su1}
\end{eqnarray}

To use of the above equation, we redefine parameters of BIon for
mth order of Lovelock gravity:

\begin{eqnarray}
&&\bar{f}=D'^{1/2}\bar{H'}^{-1/2}f',\nonumber
\\
&&f'=1-\frac{r_{0}^{4}}{[r^{[4(\frac{D-(2m+1)}{m})]}]},\nonumber
\\&&\bar{H'}=1+\frac{r_{0}^{4}}{[r^{[4(\frac{D-(2m+1)}{m})]}]}sinh^{2}\alpha \nonumber
\\&&D'^{-1}=cos^{2}\varepsilon' + H'^{-1}sin^{2}\varepsilon' \nonumber
\\&&cos\varepsilon' =\frac{1}{\sqrt{1+\frac{\beta^{2}}{[l_{2}^{[4(\frac{D-(2m+1)}{m})]}]}}}
\label{su2}
\end{eqnarray}

For D=4 and m=1, this equation converts to equation (\ref{Q3}).
For higher orders of m, the Brown-York energy evolves  faster than
lower orders (see Figure1(Right)). This is because that by
increasing the number of dimensions, more channels for flowing
energy into our universe creates which leads to emergence of
higher order terms in gravity. In these conditions, the effect of
Lovelock gravity on the system becomes more and consequently, the
energy of system increases. By considering the evolution of this
energy, we can predict some phenomenological events that occur as
due to interaction of branes in our four dimensional universe.

%%%%%%%%%%%%%%%%%%%%%%%%%%%%%%%%%%%%%%%%%%%%%%%%%%%%%%%%%%%%%%%%%%%%%%%%%%%%%%%%%%%%%%%%%%%%%%%%%%%%%%%%%%%%%
%%%%%%%%%%%%%%%%%%%%%%%% Figure 1%%%%%%%%%%%%%%%%%%%%%%%%%%%%%%%%%%%%%%%%%%%%%%%%%%%%%%%%%%%%%%%%%%%%%%%%%%%%
\begin{figure*}[thbp]
\begin{center}
\begin{tabular}{rl}
\includegraphics[width=5.6cm]{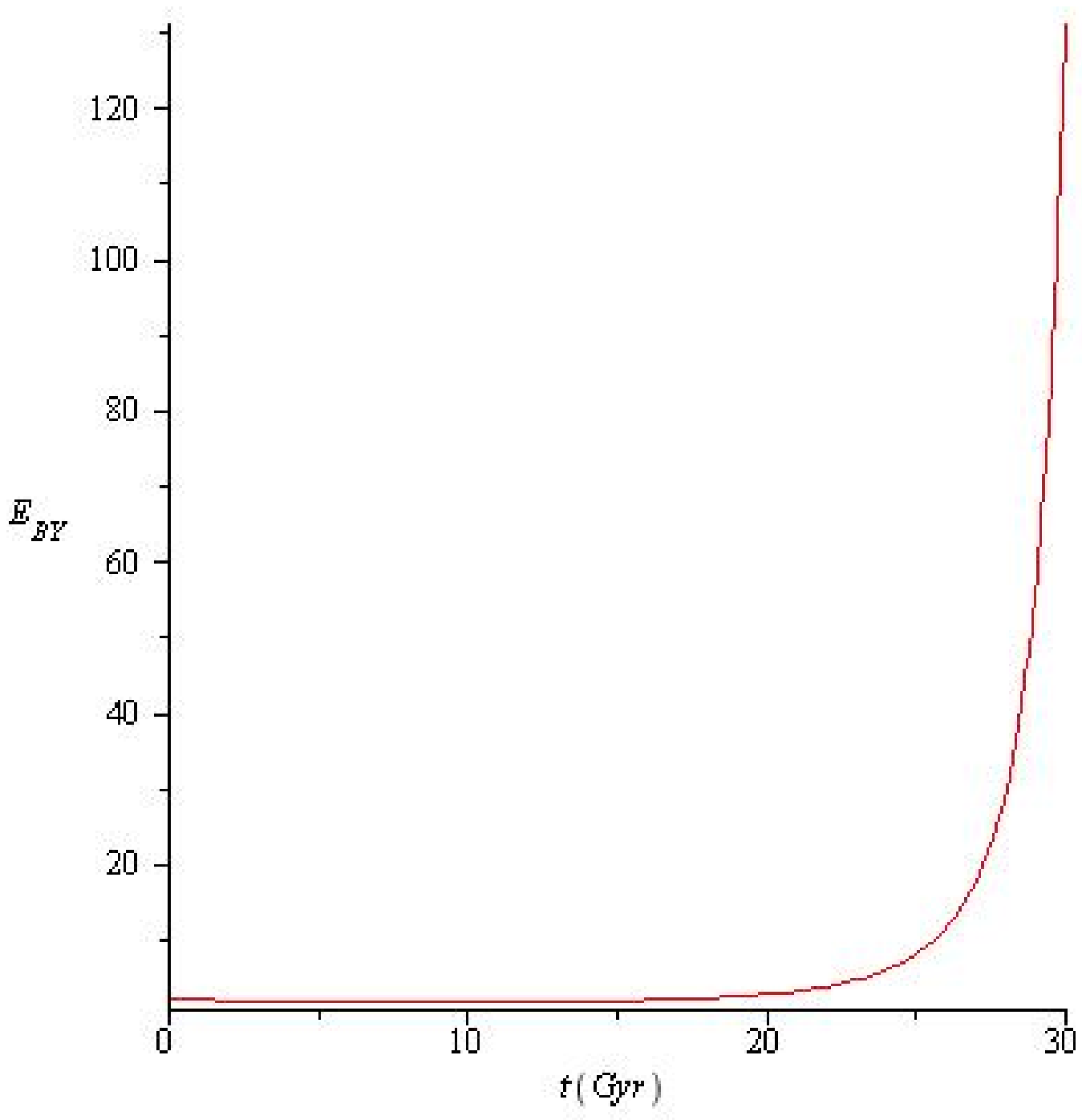}&
\includegraphics[width=5.6cm]{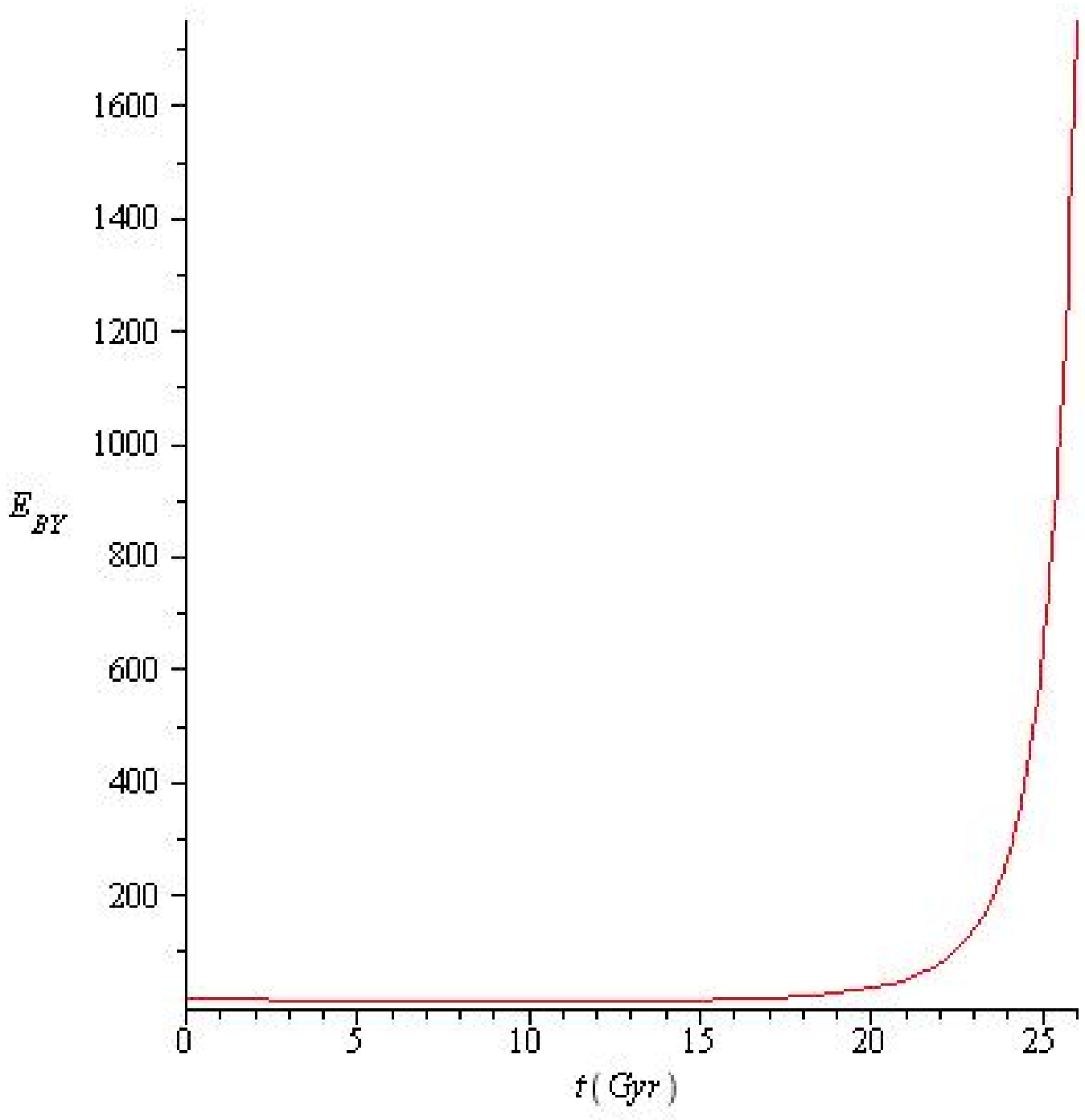}\\
\end{tabular}
\end{center}
\caption{ (left)  The  Brown-York energy for expansion branch  of
universe  with D=4 and m=1 as a function of the t where t is the
age of universe. In this plot, we choose $t_{s}=33 Gyr$,
$T_{M2}=100.$ and $l_{s}=0.1$.  (right)  In the right panel by
taking the same values of the model  parameters with  $m=3$.}
\end{figure*}

%%%%%%%%%%%%%%%%%%%%%%%%%%%%%%%%%%%%%%%%%%%%%%%%%%%%%%%%%%%%%%%%%%%%
%%%%%%%%%%%%%%%%%%%%%%%%%%%%%%%%%%%%%%%%%%%%%%%%%%%%%%%%%%%%%%%%%%%%%%%%%%%%%%%%%%%%%%%%%%%%%%%%%%%%%%%%%%%%%%%%%%

%%%%%%%%%%%%%%%%%%%%%%%%%%%%%%%%%%%%%%%%%%%%%%%%%%%%%%%%%% Section 3 %%%%%%%%%%%%%%%%%%%%%%%%

\section{The evolution of the Brown-York energy during the
contraction branch in BIonic system}\label{o2} By now, we have
shown that by closing M3-branes to each other, M2 dissolves in
them and nonlinear gravities like Lovelock emerge. Now, we will
show that by approaching branes to each other, the square energy
of system become negative and system transits to tachyonic phase.
To remove this state, Mp-branes compact, the sign of gravity
changes and anti-gravity emerges which leads to getting away of
branes from each other.  In these conditions, temperature of
system decreases, black hole disappears and the Brown-York energy
vanishes. Using equation (\ref{s28}) we have:

\begin{eqnarray}
&&l_{2}\rightarrow 0, l_{3}\rightarrow\infty, l_{2}'\rightarrow
\infty, l_{3}'\rightarrow\infty \Rightarrow\nonumber\\&&
l_{2}^{6}+l_{2}'^{2}l_{2}^{2}-\lambda^{2}l_{3}'^{2}<0
\nonumber\\&&
l_{3}^{6}+l_{3}'^{2}l_{3}^{2}-\lambda^{2}l_{2}'^{2}<0 \label{w1}
\end{eqnarray}

This equation shows that by closing branes, the quantity under
$\sqrt{}$ becomes negative and some tachyonic states are appeared.
To solve this problem, branes are compactified and Mp-branes
transit to Dp-branes. To show this, we use of the method in
\cite{m18} and define ${\displaystyle
<X^{10}>=\frac{r}{l_{p}^{3/2}}}$ where $l_{p}$ is the Planck
length. We can write:

\begin{eqnarray}
&& [X^{a},X^{b},X^{c}]=F^{abc} \quad [X^{a},X^{b}]=F^{ab}
 \nonumber \\
&&F_{abc}=\partial_{a} A_{bc}-\partial_{b} A_{ca}+\partial_{c}
A_{ab} \quad F_{ab}=\partial_{a} A_{b}-\partial_{b} A_{a}\nonumber \\ \nonumber \\
&& \nonumber \\
\Sigma_{a,b,c=0}^{10} \langle
F^{abc},F_{abc}\rangle&=&\Sigma_{a,b,c=0}^{10}
\langle[X^{a},X^{b},X^{c}],[X_{a},X_{b},X_{c}]\rangle\nonumber \\
&=&
- \Sigma_{a,b,c,a'b'c'=0}^{10}\varepsilon_{abcD}\varepsilon_{a'b'c'G}^{D}X^{a}X^{b}X^{c}X_{a'}X_{b'}X_{c'} \nonumber \\
&=& -
6\Sigma_{a,b,a',b'=0}^{9}\varepsilon_{ab10D}\varepsilon_{a'b'10}^{D}X^{a}X^{b}X^{10}X_{a'}X_{b'}X_{10}
\nonumber \\
&-&  6\Sigma_{a,b,c,a',b',c'=0,\neq
10}^{9}\varepsilon_{abcD}\varepsilon_{a'b'c'}^{D}X^{a}X^{b}X^{c}X_{a'}X_{b'}X_{c'}
\nonumber \\&=&
 - 6\left(\frac{R^{2}}{l_{p}^{3}}\right)\Sigma_{a,b,a',b'=0}^{9}\varepsilon_{ab10D}\varepsilon_{a'b'10}^{D}X^{a}X^{b}X_{a'}X_{b'} \nonumber \\
&-&
6\Sigma_{a,b,c,a',b',c'=0,\neq10}^{9}\varepsilon_{abcD}\varepsilon_{a'b'c'}^{D}X^{a}X^{b}X^{c}X_{a'}X_{b'}X_{c'}  \nonumber \\
&=& - 6\left(\frac{R^{2}}{l_{p}^{3}}\right)\Sigma_{a,b=0}^{9}[X^{a},X^{b}]^{2} \nonumber \\
&-&  6\Sigma_{a,b,c,a',b',c'=0,\neq10}^{9}\varepsilon_{abcD}\varepsilon_{a'b'c'}^{D}X^{a}X^{b}X^{c}X_{a'}X_{b'}X_{c'} \nonumber \\
&=& -
6\left(\frac{R^{2}}{l_{p}^{3}}\right)\Sigma_{a,b=0}^{9}F^{ab}F_{ab}
+ E_{Extra} \nonumber\\&&
\nonumber\\&&\partial_{a}X^{i}\partial_{a}X^{i}\sum
(X^{j})^{2}=\Sigma_{a,i,j=0}^{10}\langle[X^{a},X^{i},X^{j}],[X_{a},X_{i},X_{j}]\rangle\Rightarrow
\nonumber\\&&
-6\left(\frac{R^{2}}{l_{p}^{3}}\right)\Sigma_{i,j=0}^{9}[X^{a},X^{j}][X_{a},X_{j}]=-6\left(\frac{R^{2}}{l_{p}^{3}}\right)
\partial_{a}X^{i}\partial_{a}X^{i}\Rightarrow
F(X)=\sum
(X^{j})^{2}=1\nonumber\\
&&\partial_{a}\partial_{b}
X^{i}\partial_{a}\partial_{b}X^{i}=\Sigma_{a,b,i=0}^{10}\langle[X^{a},X^{b},X^{i}],[X_{a},X_{b},X_{i}]\rangle
\Rightarrow \nonumber\\&&
-6\left(\frac{R^{2}}{l_{p}^{3}}\right)\Sigma_{i,b=0}^{9}[X^{b},X^{i}][X_{b},X_{i}]=-6\left(\frac{R^{2}}{l_{p}^{3}}\right)\partial_{a}
X^{i}\partial_{a}X^{i}\label{w2}
\end{eqnarray}

This equation shows that by compactification of branes, two form
fields in eleven dimensional space-time converts to one form field
 and the sign of nonlinear terms in action change. Using equations (\ref{s21}) and (\ref{w2}),
we can obtain the relation between noncompact and compact gravity
:

\begin{eqnarray}
&&A_{b}=e_{b}\quad F_{ab}=\partial_{a} e_{b}-\partial_{b} e_{a}\quad \kappa^{a}\smallskip_{b}=\partial^{a} e_{b}\nonumber \\
&&\Sigma_{\rho,\sigma,\mu,\nu=0}^{10}R^{\rho\sigma}_{\mu\nu}=\Sigma_{\rho,\sigma,\mu,\nu,\lambda=0}^{10}\langle
F^{\rho\sigma}\smallskip_{\lambda},F^{\lambda}\smallskip_{\mu\nu}\rangle=\Sigma_{\rho,\sigma,\mu,\nu,\lambda=0}^{10}
\langle[X^{\rho},X^{\sigma},X_{\lambda}],[X^{\lambda},X_{\mu},X_{\nu}]\rangle\Rightarrow
\nonumber\\&&
-6\left(\frac{R^{2}}{l_{p}^{3}}\right)\Sigma_{\rho,\sigma,\mu,\nu=0}^{9}[X^{\rho},X^{\sigma}][X_{\mu},X_{\nu}]=
-6\left(\frac{R^{2}}{l_{p}^{3}}\right)
\Sigma_{\rho,\sigma,\mu,\nu=0}^{9}F^{\rho\sigma}F_{\mu,\nu}=\nonumber\\&&-6(\frac{R^{2}}{l_{p}^{3}})
\Sigma_{\rho,\sigma,\mu,\nu=0}^{9}\delta^{\rho\sigma}_{\rho'\sigma'}\delta_{\mu\nu}^{\mu'\nu'}\partial^{\rho'}
e^{\sigma'}\partial_{\mu'}
e_{\nu'}=-6\left(\frac{R^{2}}{l_{p}^{3}}\right)
\Sigma_{\rho,\sigma,\mu,\nu=0}^{9}\delta^{\rho\sigma}_{\rho'\sigma'}\delta_{\mu\nu}^{\mu'\nu'}\kappa^{\rho'\sigma'}
\kappa_{\mu'\nu'}\nonumber\\&&
\Sigma_{\rho,\sigma,\mu,\nu=0}^{10}R=-6\left(\frac{R^{2}}{l_{p}^{3}}\right)
\Sigma_{\rho,\sigma,\mu,\nu=0}^{9}\delta^{\mu\nu}_{\rho'\sigma'}\delta_{\mu\nu}^{\mu'\nu'}\kappa^{\rho'\sigma'}
\kappa_{\mu'\nu'}\label{w3}
\end{eqnarray}

Above equation helps us to show that the sign of  gravity changes:

\begin{eqnarray}
&&\sum_{n=1}^{p}m_{g}^{2n}\delta^{\rho_{1}\sigma_{1}...\rho_{n}\sigma_{n}}_{\mu_{1}\nu_{1}...\mu_{n}\nu_{n}}
R_{\rho_{1}\sigma_{1}}^{\mu_{1}\nu_{1}}...R_{\rho_{n}\sigma_{n}}^{\mu_{n}\nu_{n}}\Rightarrow
-\sum_{n=1}^{p}m_{g}^{2n}\left(\frac{R^{2n}}{l_{p}^{3n}}\right)
\delta^{\rho_{1}\sigma_{1}...\rho_{n}\sigma_{n}}_{\mu_{1}\nu_{1}...\mu_{n}\nu_{n}}\kappa^{\mu_{1}}_{\rho_{1}}
\kappa^{\nu_{1}}_{\sigma_{1}}
..\kappa^{\mu_{n}}_{\rho_{n}}\kappa^{\nu_{n}}_{\sigma_{n}}\nonumber\\&&
\sum_{n=1}^{p}m_{g}^{2n}R^{n}\Rightarrow
-\sum_{n=1}^{p}m_{g}^{2n}\left(\frac{R^{2n}}{l_{p}^{3n}}\right)
\kappa^{\mu_{1}}_{\rho_{1}} \kappa^{\rho_{1}}_{\mu_{1}}
..\kappa^{\mu_{n}}_{\rho_{n}}
\kappa^{\rho_{n}}_{\mu_{n}}\label{w4}
\end{eqnarray}

Substituting relations in equation (\ref{w4}) in equation
(\ref{s27}), we obtain:

\begin{eqnarray}
S_{Mp} &=& -(T_{Mp})\int dt \int d^{p}\sigma
\Biggl[\sqrt{-g}\Bigl(\sum_{n=1}^{p}\beta_{n}[-F(X)^{n}(1-m_{g}^{2})^{n}
\left(\frac{R^{2n}}{l_{p}^{3n}}\right) \kappa^{\mu_{1}}_{\rho_{1}}
\kappa^{\rho_{1}}_{\mu_{1}} ..\kappa^{\mu_{n}}_{\rho_{n}}
\kappa^{\rho_{n}}_{\mu_{n}}-\nonumber\\&&
\sum_{n=1}^{p}m_{g}^{2n}\lambda^{2n}\left(\frac{R^{2n}}{l_{p}^{3n}}\right)
\delta^{\rho_{1}\sigma_{1}...\rho_{n}\sigma_{n}}_{\mu_{1}\nu_{1}...\mu_{n}\nu_{n}}\kappa^{\mu_{1}}_{\rho_{1}}
\kappa^{\nu_{1}}_{\sigma_{1}}
..\kappa^{\mu_{n}}_{\rho_{n}}\kappa^{\nu_{n}}_{\sigma_{n}}]\Bigr)\Biggr]
\label{w5}
\end{eqnarray}

This equation indicates that when Mp-branes are compactified,
nonlinear gravity terms changes to other type of non-linear
gravity terms with opposite sign. This can be a signature of
anti-gravity. As due to the emergence of anti-gravity, M3-branes
get away from each other and one M2-branes is produced between
them. In our model, branes and anti-branes live on the M3-branes
and thus, by closing towards each other, gravity changes to
anti-gravity and prevents from approaching and disappearing
branes. Using the relations in equation (\ref{w2}), we can show
that the action (\ref{s14}) converts to following action:

\begin{eqnarray}
 &&S=-\frac{T_{Mp}}{2}\int d^{p+1}x \sum_{n=1}^{p}\beta_{n} \tilde{\chi}^{\mu_{0}}_{[\mu_{0}}\tilde{\chi}^{\mu_{1}}_{\mu_{1}}
 ...\tilde{\chi}^{\mu_{n}}_{\mu_{n}]}\nonumber\\&&\tilde{\chi}\equiv STr \Bigg(-det(P_{ab}[E_{mn}
E_{mi}(Q^{-1}+\delta)^{ij}E_{jn}+\lambda
F_{ab}])det(Q^{i}_{j})\Bigg)^{1/2}~~ \nonumber\\&&
   E_{mn} = G_{mn} + B_{mn}, \qquad  Q^{i}_{j} = \delta^{i}_{j} + i\lambda[X^{j},X^{k}]E_{kj}\nonumber\\&&G_{ab}=\eta_{ab}+\partial_{a}X^{i}\partial_{b}X^{i} \label{r2}\label{w6}
\end{eqnarray}

As can be seen from above equation, by compactifying M-theory,
Mp-branes transit to Dp-branes. Similar to previous section, we
assume that universes are placed on D3-branes which are connected
by a D2-brane. To consider the evolution of branes, we assume  the
length of D2 between two D3 is $l_{2}$ and the length of each D3
be $l_{3}$ and obtain the relevant action for the system of D3-D2:

\begin{eqnarray}
&&  A^{ab}\rightarrow l_{3} \quad X^{i}\rightarrow l_{2} \quad
\text{for D3-brane}
\nonumber\\&& A^{ab}\rightarrow l_{2} \quad X^{i}\rightarrow l_{3} \quad \text{for D2-brane}\nonumber\\
&&  S_{tot}=S_{D3}+S_{D2}+S_{anti-D3}=\nonumber\\&&-2T_{D3} \int
d^{6}\sigma \Sigma_{n=1}^{3}
\beta_{n}(-l_{2}^{4}+l_{2}'^{2}l_{2}^{2}+\lambda^{2}l_{3}'^{2})^{\frac{n}{2}}\nonumber\\&&-2T_{D2}
\int d^{3}\sigma \Sigma_{n=1}^{2}\beta_{n}
(-l_{3}^{4}+l_{3}'^{2}l_{3}^{2}+\lambda^{2}l_{2}'^{2})^{\frac{n}{2}}\simeq\nonumber\\&&-2T_{D3}
\int dt l_{3}^{3} \Sigma_{n=1}^{3}
\beta_{n}(-l_{2}^{4}+l_{2}'^{2}l_{2}^{2}+\lambda^{2}l_{3}'^{2})^{\frac{n}{2}}\nonumber\\&&-2T_{D2}
\int dt l_{2}^{2} \Sigma_{n=1}^{3}\beta_{n}
(-l_{3}^{4}+l_{3}'^{2}l_{3}^{2}+\lambda^{2}l_{2}'^{2})^{\frac{n}{2}}\label{w7}
\end{eqnarray}

The equation of motion for this equation is:

\begin{eqnarray}
&&-2T_{D3} \Sigma_{n=1}^{3} \beta_{n}[ l_{3}^{3}(
l_{2}'l_{2}^{2})(-l_{2}^{4}+l_{2}'^{2}l_{2}^{2}+\lambda^{2}l_{3}'^{2})^{\frac{n}{2}-1}]'\nonumber\\&&
+2T_{D2} \Sigma_{n=1}^{2} \beta_{n}[(\lambda^{2} l_{2}^{2}
l_{2}'))(-l_{3}^{4}+l_{3}'^{2}l_{3}^{2}+\lambda^{2}l_{2}'^{2})^{\frac{n}{2}-1}]'\nonumber\\&&
+2T_{D3} \Sigma_{n=1}^{3} \beta_{n}[ l_{3}^{3}(
l_{2}'^{2}l_{2}+2l_{2}^{3})(-l_{2}^{4}+l_{2}'^{2}l_{2}^{2}+\lambda^{2}l_{3}'^{2})^{\frac{n}{2}-1}]\nonumber\\&&
-2T_{D2} \Sigma_{n=1}^{2} \beta_{n}[(
l_{2}))(-l_{3}^{4}+l_{3}'^{2}l_{3}^{2}+\lambda^{2}l_{2}'^{2})^{\frac{n}{2}}]=0
\label{w8}
\end{eqnarray}

\begin{eqnarray}
&&-2T_{D2} \Sigma_{n=1}^{2} \beta_{n}[ l_{2}^{2}(
l_{3}'l_{3}^{2})(-l_{3}^{4}+l_{3}'^{2}l_{3}^{2}+\lambda^{2}l_{2}'^{2})^{\frac{n}{2}-1}]'\nonumber\\&&
+2T_{D3} \Sigma_{n=1}^{3} \beta_{n}[(\lambda^{2} l_{3}^{3}
l_{3}'))(-l_{2}^{4}+l_{2}'^{2}l_{2}^{2}+\lambda^{2}l_{3}'^{2})^{\frac{n}{2}-1}]'\nonumber\\&&
+2T_{D2} \Sigma_{n=1}^{2} \beta_{n}[ l_{2}^{2}(
l_{3}'^{2}l_{3}+2l_{3}^{3})(-l_{3}^{4}+l_{3}'^{2}l_{3}^{2}+\lambda^{2}l_{2}'^{2})^{\frac{n}{2}-1}]\nonumber\\&&
-2T_{D3} \Sigma_{n=1}^{5} \beta_{n}[(3
l_{3}^{2}))(-l_{2}^{4}+l_{2}'^{2}l_{2}^{2}+\lambda^{2}l_{3}'^{2})^{\frac{n}{2}}]=0
\label{w9}
\end{eqnarray}

We can obtain the following approximate solutions for this
equation:

\begin{eqnarray}
&&l_{3} \simeq \sum_{n=1}^{3}\beta_{n}\frac{n}{\lambda^{n}
T_{3}^{6n}}
(\frac{t}{t_{s}}-1)^{-n}e^{-n\lambda\frac{t}{t_{s}}}\nonumber\\&&
l_{2} \simeq \sum_{n=1}^{2}\beta_{n}\frac{1}{\lambda^{n}
T_{2}^{3n}}(\frac{t}{t_{s}}-1)^{n}(1-e^{-n\lambda[\frac{t}{t_{s}}-1]})
\label{w10}
\end{eqnarray}

This equation shows that at $t=t_{s}$, the separation distance
($l_{2}$) between two branes is zero and the size of D3 branes is
infinite, while by passing time, branes get away from each other,
$l_{2}\rightarrow\infty$ and $l_{3}\rightarrow 0$. This means that
branes and anti-branes get away from each other and connect by a
D2-brane. Similar to previous section, using equations (\ref{w7}
and \ref{w10}), we derive the effective potential between branes
and anti-branes:

\begin{eqnarray}
&&t\rightarrow t_{s}\Rightarrow l_{3}\simeq l_{3}'  \simeq
\frac{1}{l_{2}} \quad l_{2}' \simeq \frac{1}{l_{2}}\nonumber\\&&
V_{brane-antibrane}=2T_{M3}  l_{2}^{-5} \Sigma_{n=1}^{5}
\beta_{n}(l_{2}^{4}+1+\lambda^{2}l_{2}^{-2})^{\frac{n}{2}}\nonumber\\&&+2T_{M2}
l_{2}^{2} \Sigma_{n=1}^{3}\beta_{n}
(2l_{2}^{-4}+\lambda^{2}l_{2}^{-2})^{\frac{n}{2}}\simeq
\nonumber\\&&
b_{1}l_{2}+b_{2}l_{2}^{-3}+b_{3}l_{2}^{-5}+b_{4}l_{2}^{-6}+....
\label{w11}
\end{eqnarray}

where $b_{i}$'s are some coefficients which  depend on the tension
of branes and $\lambda$. This equation shows that by compacting
branes, order of $l_{2}^{-1}$ increases. This is because that near
the colliding point $l_{2}\rightarrow0$, the repulsive potential
is very large and shrinks to zero at infinity.  Now, we obtain
temperature of system by using the equations (\ref{t18} and
\ref{w11}):

\begin{eqnarray}
&& T=\frac{\gamma}{\pi}(\frac{w'}{w})^{1/2}\sim
\frac{\gamma}{\pi}(\frac{V'}{V})^{1/2}\sim \nonumber
\\&&\frac{\gamma}{\pi}\left(\frac{
b_{1}+b_{2}l_{2}^{-4}+b_{3}l_{2}^{-6}+b_{4}l_{2}^{-7}}{
b_{1}l_{2}+b_{2}l_{2}^{-3}+b_{3}l_{2}^{-5}+b_{4}l_{2}^{-6}}\right)^{1/2}
\label{w13}
\end{eqnarray}

Substituting  equation (\ref{t19}) in above equation , we can
obtain  the size of M2 in terms of the temperature:

\begin{eqnarray}
&& l_{2}\simeq T^{2}[\frac{6}{k\sqrt{3}}-\bar{T}^{2}]^{2}
\label{w14}
\end{eqnarray}

Similar to previous section, by definition of
$k=\frac{\sqrt{3}}{6}$, temperature of colliding point and
critical temperature become equal ($T_{\text{colliding
point}}=T_{c}$). Substituting equations
(\ref{w10},\ref{t16},\ref{t17},\ref{t18},\ref{t19}and \ref{t20})in
equation (\ref{su1}) for m=1 and D=4 and in the limit $r\sim z\sim
l_{2}\rightarrow 0$, we obtain:

\begin{eqnarray}
&&E_{BY}\sim [\sum_{n=1}^{2}\beta_{n}\frac{1}{\lambda^{n}
T_{2}^{3n}}(\frac{t}{t_{s}}-1)^{n}(1-e^{-n\lambda[\frac{t}{t_{s}}-1]})]+\Big(\frac{l_{2,0}^{4}}{[\sum_{n=1}^{2}\beta_{n}\frac{1}{\lambda^{n}
T_{2}^{3n}}(\frac{t}{t_{s}}-1)^{n}(1-e^{-n\lambda[\frac{t}{t_{s}}-1]})
]^{3}}\times\nonumber\\&&
[\frac{1}{1+\frac{\beta^{2}}{[\sum_{n=1}^{2}\beta_{n}\frac{1}{\lambda^{n}
T_{2}^{3n}}(\frac{t}{t_{s}}-1)^{n}(1-e^{-n\lambda[\frac{t}{t_{s}}-1]})]^{4}}}
+
[1+\frac{l_{2,0}^{4}}{[\sum_{n=1}^{2}\beta_{n}\frac{1}{\lambda^{n}
T_{2}^{3n}}(\frac{t}{t_{s}}-1)^{n}(1-e^{-n\lambda[\frac{t}{t_{s}}-1]})]^{4}}
[1-\nonumber\\&&(\frac{3\sqrt{3}-\bar{T}^{4}\sqrt{1+\frac{k^{2}}{[\sum_{n=1}^{2}\beta_{n}\frac{1}{\lambda^{n}
T_{2}^{3n}}(\frac{t}{t_{s}}-1)^{n}(1-e^{-n\lambda[\frac{t}{t_{s}}-1]})
]^{4}}}
-\frac{\sqrt{3}}{6}\bar{T}^{8}(1+\frac{k^{2}}{[\sum_{n=1}^{2}\beta_{n}\frac{1}{\lambda^{n}
T_{2}^{3n}}(\frac{t}{t_{s}}-1)^{n}(1-e^{-n\lambda[\frac{t}{t_{s}}-1]})]^{4}})}
{2\bar{T}^{4}\sqrt{1+\frac{k^{2}}{[\sum_{n=1}^{2}\beta_{n}\frac{1}{\lambda^{n}
T_{2}^{3n}}(\frac{t}{t_{s}}-1)^{n}(1-e^{-n\lambda[\frac{t}{t_{s}}-1]})
]^{4}}}})^{2}]]^{-1}\times
\nonumber\\&&[1-\frac{1}{1+\frac{\beta^{2}}{[\sum_{n=1}^{2}\beta_{n}\frac{1}{\lambda^{n}
T_{2}^{3n}}(\frac{t}{t_{s}}-1)^{n}(1-e^{-n\lambda[\frac{t}{t_{s}}-1]})
]^{4}}}]]^{-1/2}
[1+\frac{l_{2,0}^{4}}{[\sum_{n=1}^{2}\beta_{n}\frac{1}{\lambda^{n}
T_{2}^{3n}}(\frac{t}{t_{s}}-1)^{n}(1-e^{-n\lambda[\frac{t}{t_{s}}-1]})]^{4}}\times
\nonumber\\&&[(\frac{3\sqrt{3}-\bar{T}^{4}\sqrt{1+\frac{k^{2}}{[\sum_{n=1}^{2}\beta_{n}\frac{1}{\lambda^{n}
T_{2}^{3n}}(\frac{t}{t_{s}}-1)^{n}(1-e^{-n\lambda[\frac{t}{t_{s}}-1]})]^{4}}}
-\frac{\sqrt{3}}{6}\bar{T}^{8}(1+\frac{k^{2}}{[\sum_{n=1}^{2}\beta_{n}\frac{1}{\lambda^{n}
T_{2}^{3n}}(\frac{t}{t_{s}}-1)^{n}(1-e^{-n\lambda[\frac{t}{t_{s}}-1]})
]^{4}})}{2\bar{T}^{4}\sqrt{1+\frac{k^{2}}{[\sum_{n=1}^{2}\beta_{n}\frac{1}{\lambda^{n}
T_{2}^{3n}}(\frac{t}{t_{s}}-1)^{n}(1-e^{-n\lambda[\frac{t}{t_{s}}-1]})
]^{4}}}})^{2}]]^{-1/2}\Big)\label{Q5}\nonumber \\
\end{eqnarray}

This equation shows that by passing time and increasing the
separation distance between two branes ($l_{2}$), the Brown-York
energy decreases and shrinks to zero at large values of time (See
Figure 2(Left).). This is because that by getting away of
M3-branes, the nonlinear gravity like the  Lovelock gravity
disappears and temperature and also the gravitational energy of
system decreases. Using equation (\ref{su1}), we can obtain the
Brown-York energy for mth order of Lovelock gravity for a system
with D=3m+1

\begin{eqnarray}
&&\overline{E}_{BY}=r^{m}[D_{0}(\frac{E_{BY}}{r})^{2m-1}+...+D_{s}(\frac{E_{BY}}{r})^{2m-2s-1}(1-D^{1/2}\bar{H}^{-1/2}f)^{s}+...D_{m-1}(\frac{E_{BY}}{r})^{m-1}
(1-\sqrt{D^{1/2}\bar{H}^{-1/2}f})] \label{su1} \nonumber \\
\end{eqnarray}

For m=1, above equation reduces to equation (\ref{Q5}). This
equation shows that by increasing number of dimensions  the effect
of Lovelock gravity increases and consequently, more changes can
be observed in energy of system (See Figure2(Right)). This is
because that by increasing the number of dimensions, there exists
many channels for flowing energy from extra dimensions and other
universes to our own universe which leads to the emergence of
higher order terms in gravity.

%%%%%%%%%%%%%%%%%%%%%%%%%%%%%%%%%%%%%%%%%%%%%%%%%%%%%%%%%%%%%%%%%%%%%%%%%%%%%%%%%%%%%%%%%%%%%%%%%%%%%%%%%%%%%
%%%%%%%%%%%%%%%%%%%%%%%% Figure 1%%%%%%%%%%%%%%%%%%%%%%%%%%%%%%%%%%%%%%%%%%%%%%%%%%%%%%%%%%%%%%%%%%%%%%%%%%%%
\begin{figure*}[thbp]
\begin{center}
\begin{tabular}{rl}
\includegraphics[width=5.6cm]{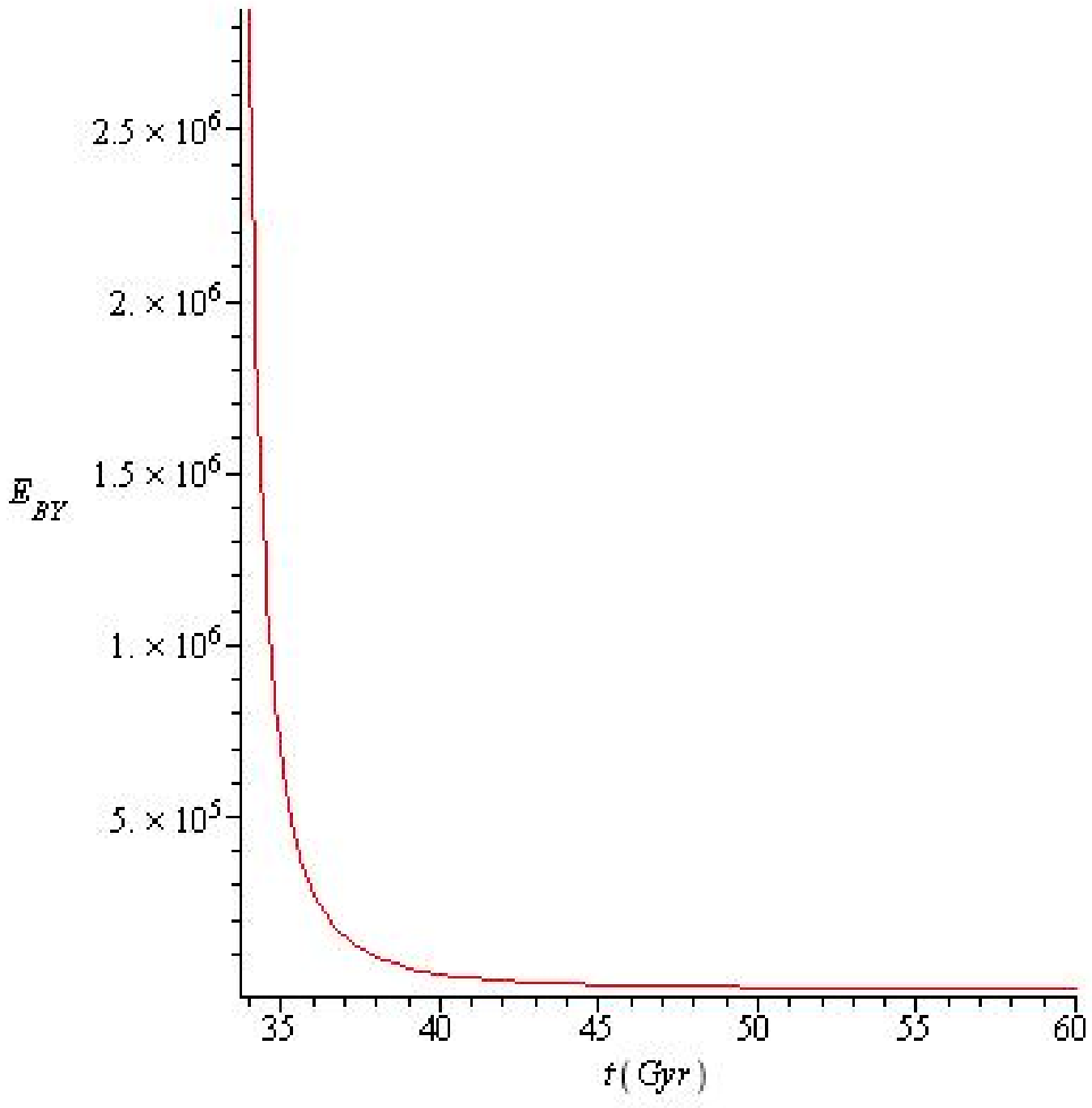}&
\includegraphics[width=5.6cm]{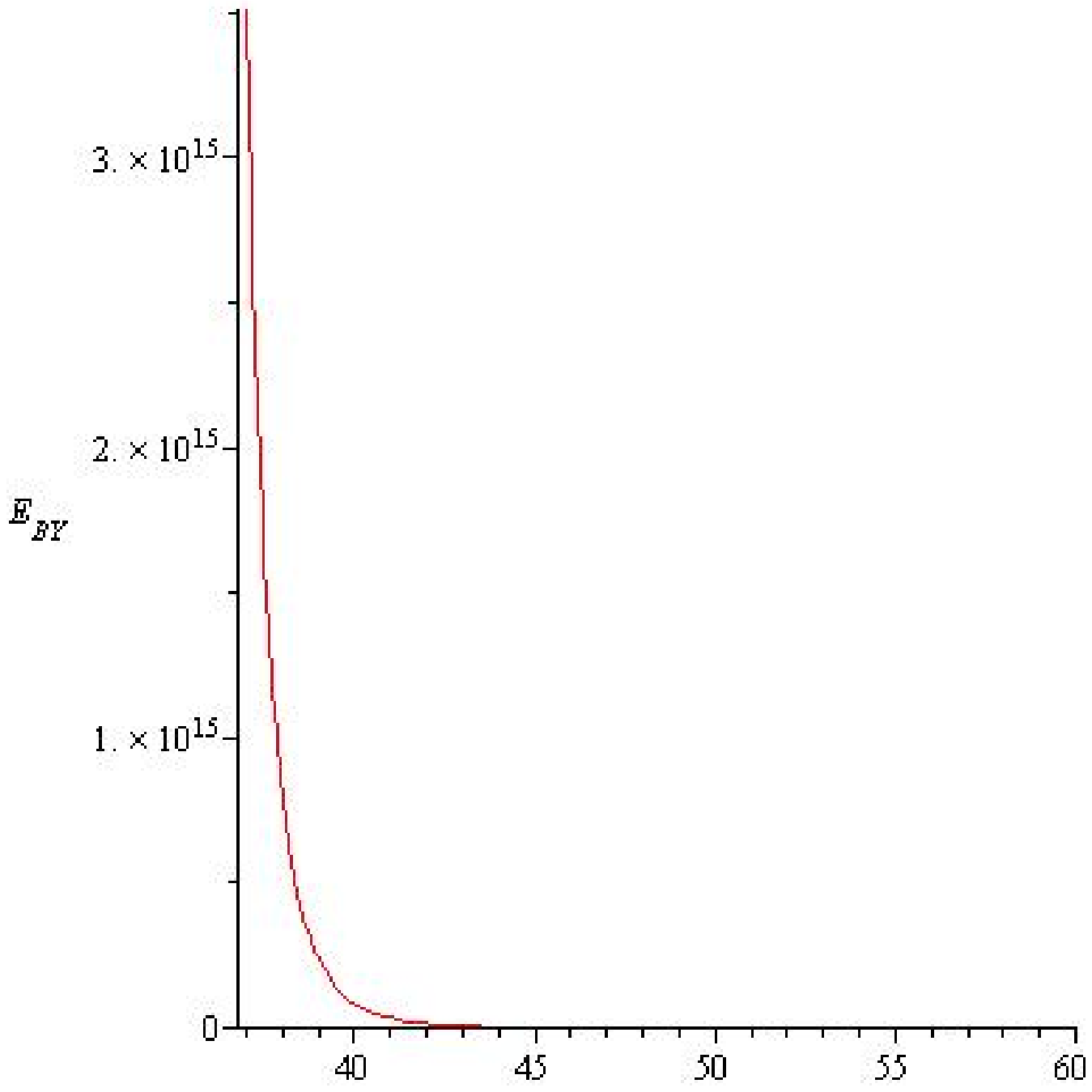}\\
\end{tabular}
\end{center}
\caption{ (left) The  Brown-York energy for contraction branch  of
universe with D=4 and m=1 as a function of the t where t is the
age of universe. In this plot, we choose $t_{s}=33 Gyr$,
$T_{M2}=100.$ and $l_{s}=0.1$.  (right)  In the right panel by
taking the same values of the model  parameters with $m=3$.}
\end{figure*}

%%%%%%%%%%%%%%%%%%%%%%%%%%%%%%%%%%%%%%%%%%%%%%%%%%%%%%%%%%%%%%%%%%%%
%%%%%%%%%%%%%%%%%%%%%%%%%%%%%%%%%%%%%%%%%%%%%%%%%%%%%%%%%%%%%%%%%%%%%%%%%%%%%%%%%%%%%%%%%%%%%%%%%%%%%%%%%%%%%%%%%%

%%%%%%%%%%%%%%%%%%%%%%%%%%%%%%%%%%%%%%%%%%%%%%%%%%%% Section 4 %%%%%%%%%%%%%%%%%%%%%%%%%%%%%%%%%

\section{Summary and Discussion} \label{sum}
In this research, we have used the idea of \cite{s5} for
measuring the gravitational energy in Lovelock gravity for
considering the evolution of BIonic system. In our model, first
M0-branes join each other and form a system of one M3, one
anti-M3 and an M2-brane. Brane lives on an M3, anti-brane is
placed on an anti-M3 and M2 connects them. M2 plays the role of
wormhole between branes and is the main cause of gravitational
energy in BIonic system. By dissolving M2 in M3-branes, Lovelock
gravity is produced and M3 expands. By closing branes to each
other, wormhole becomes thermal and transits to black hole. During this epoch, the Brown-York increases and shrinks to large values. By approaching branes, the square energy of M2-M3 system becomes negative and some tachyonic states are created. To remove these states, M2 and M3-branes compact, gravity converts to anti-gravity and branes get away from each other. We have calculated the potential between branes and anti-branes and find that it is in agreement with previous predictions. This means that usual
potential between branes may be produced by dissolving M2-branes
in M3-branes which branes live on it. On the other hand, by
closing branes and disappearing M2, Lovelock gravity emerges.
Thus, the Lovelock gravity may be the main cause of the wormhole
and gravitational energy between branes and anti-branes. During
this era, the Brown-York decreases and shrinks to zero.

%%%%%%%%%%%%%%%%%%%%%%%%%%%%%%%%%%%%%%%%%%%%%%%%%%%%%%%%%%%%%%%%%%%%%%%%%%%%%%%%%%%%%%%%%

\section*{Acknowledgments}
\noindent  Alireza Sepehri acknowledges the Research Institute for
Astronomy and Astrophysics of Maragha, Iran for financial support
during this work. Also, he would like to thank of Sumanta
Chackraborty for useful discussions.

 \end{document}